\documentclass[reprint,superscriptaddress,
amsmath,amssymb,aps,
]{revtex4-2}
\usepackage{graphicx}% Include figure files
\usepackage{dcolumn}% Align table columns on decimal point
\usepackage{bm}% bold math
\usepackage{amsfonts}
\usepackage{xcolor}
\usepackage{natbib}

\DeclareMathOperator*{\SumInt}{%
\mathchoice%
  {\ooalign{$\displaystyle\sum$\cr\hidewidth$\displaystyle\int$\hidewidth\cr}}
  {\ooalign{\raisebox{.14\height}{\scalebox{.7}{$\textstyle\sum$}}\cr\hidewidth$\textstyle\int$\hidewidth\cr}}
  {\ooalign{\raisebox{.2\height}{\scalebox{.6}{$\scriptstyle\sum$}}\cr$\scriptstyle\int$\cr}}
  {\ooalign{\raisebox{.2\height}{\scalebox{.6}{$\scriptstyle\sum$}}\cr$\scriptstyle\int$\cr}}
}

\begin{document}

\title{Phase boundaries promote chemical reactions through localized fluxes}

\author{Alexandra Shelest}
\affiliation{Institute of Physics, \'Ecole Polytechnique F\'ed\'erale de Lausanne (EPFL), 1015 Lausanne, Switzerland}
\author{Hugo Le Roy} 
\affiliation{Institute of Physics, \'Ecole Polytechnique F\'ed\'erale de Lausanne (EPFL), 1015 Lausanne, Switzerland}
\author{Daniel M. Busiello}
\affiliation{Max Planck Institute for the Physics of Complex Systems, 01187 Dresden, Germany}
\author{Paolo De Los Rios}
\affiliation{Institute of Physics, \'Ecole Polytechnique F\'ed\'erale de Lausanne (EPFL), 1015 Lausanne, Switzerland}
\affiliation{Institute of Bioengineering, \'Ecole Polytechnique F\'ed\'erale de Lausanne (EPFL), 1015 Lausanne, Switzerland}

\date{\today}

\begin{abstract}
One of the hypothesized functions of biomolecular condensates is to act as chemical reactors, where chemical reactions can be modulated, \textit{i.e.} accelerated or slowed down, while substrate molecules enter and products exit from the condensate. Likewise, the components themselves that take part in the architectural integrity of condensates might be modified by active (energy consuming, non-equilibrium) processes, \textit{e.g.} by ATPase chaperones or by kinases and phosphatases. In this work, we study how the presence of spatial inhomogeneities, such as in the case of liquid-liquid phase separation, affects active chemical reactions and results in the presence of directional flows of matter, which are one of the hallmarks of non-equilibirum processes. We establish the minimal conditions for the existence of such spatial currents, and we furthermore find that these fluxes are maximal at the condensate interface. These results propose that some condensates might be most efficient as chemical factories due to their interfaces rather than their volumes, and could suggest a possible biological reason for the the observed abundance of small non-fusing condensates inside the cell, thus maximizing their surface and the associated fluxes. 
\end{abstract}

\maketitle
\section{Introduction} \label{sec: intro} 
Biomolecular condensates are membraneless cellular organelles that can specifically control their own composition, being enriched in some components and depleted in others. This barrierless partitioning offers, besides other properties, the possibility to concentrate specific enzymes and their substrates, and to potentially accelerate the rates of the reactions they catalyze, possibly resulting in fluxes of molecules both to and from condensates. Furthermore, the molecules that are structurally responsible for the condensate formation might themselves undergo transformations, driven by chemical reactions, that tune their conformational states and their interactions, consequently modulating the condensate properties. For example, \citet{Cisse} found that ruvBL, an AAA+ ATPase, is involved in the size control of Synphilin1 condensates and \citet{parABS} showed that the ATPase ParA is actively involved in preventing the fusion of parABS condensates. Another typical hallmark of active processes, the presence of fluxes, has been observed in the nucleolus, which is instrumental to the assembly of ribosomes from rRNA \citep{nucleolus}. Directional mRNA fluxes have also been measured in nuclear speckles \citep{wang2018intronless, ishov2020coordination}. There are several further systems where the presence of fluxes is hypothesized because mRNA must be sequentially modified while transiting from one condensate to another \citep{hondele}, \textit{e.g.} between P-bodies and stress granules \citep{Kedersha}. The presence of fluxes dramatically underscores the necessity of non-equilibrium driving for the molecular transformations taking place in condensates.

Physically, condensates are often modelled in the framework of liquid-liquid phase separation, according to the Flory-Huggins theory (\citep{Flory42}, \citep{Huggins42}, \citep{ActiveEmulsions}). This description is only valid at equilibrium and must be extended to account for active effects. In most works (\textit{e.g.}, \citep{ActiveEmulsions, Kirschbaum_2021}), the kinetics is derived close to equilibrium from linearized fluxes which depend on the chemical potentials of the species, calculated from the free energy, and on the chemical potential of the fuel. 

A number of studies also investigates the interplay between active condensate maintenence and the presence of fluxes \citep{ActiveEmulsions, Bauermann, Kirschbaum_2021, osmanovic2022chemical, Zwicker_2022}, yet a clear analysis of the conditions that are necessary for fluxes to be present, and of their spatial structure, is still lacking. 
This is precisely what we address in this work.

In Section~\ref{sec: intro model} we describe in detail our mathematical framework, where we derive and extend the customary free-energy approach from a microscopic perspective. In Section~\ref{sec: obtain fluxes}, we show that the simple presence of activity is not sufficient for the presence of fluxes, and we thus establish the minimal set of conditions that must be satisfied in a cell (or chemical system) to maintain steady-state spatial fluxes in the presence of an interface between two phases. In Section~\ref{sec: max fluxes}, we study how these fluxes depend on space and on some of the salient parameters of the system. We find that diffusive fluxes are always maximal close to the interface,
decreasing exponentially away from it, with the decay governed by the reaction-diffusion length. As a consequence, depending on this length and on the condensate size, the entire condensate, or just a region close to its surface, contributes to the fluxes, with relevant consequences on the current view of these membraneless organelles as chemical factories.

\section{Derivation of the formalism} \label{sec: intro model}

We model a system with a fixed number of particles, $N$, where each particle can be in any of $K$ chemical states. 
The different chemical states can correspond to, \textit{e.g.}, different protein conformations that change the way the particles interact, and hence their preferences for either the condensate or the dilute phase. We assume that the diffusion constant $D$ does not depend on the chemical state of the particle (this condition can be relaxed without loss of generality of our results). \\
The total energy of the system comprises a contribution from external potentials, $V(\vec{x}_i,\sigma_i)$, and a part due to inter-particle interactions, $U(\vec{x}_i,\vec{x}_j,\sigma_i,\sigma_j)$, both depending on the positions and states of the particles: 
\begin{equation} \label{energies}
E(\vec{x},\vec{\sigma}) = \sum_{i=1}^N V(\vec{x}_i,\sigma_i) + \sum_{i=1}^{N-1} \sum_{j=i+1}^N U(\vec{x}_i,\vec{x}_j,\sigma_i,\sigma_j)
\end{equation}
where $\vec{x} = (\vec{x}_1, ..., \vec{x}_N)$ and $\vec{\sigma} = (\sigma_1, ..., \sigma_N)$ denote the ensemble of particle positions and states, respectively, and for the sake of simplicity, we are restricting the analysis to one- and two-body interactions.

We describe the system through the probability distribution $P(\vec{x},\vec{\sigma},t)$, starting from the full $N$-particle Fokker-Planck equation \citep{Archer2004}, which is a known approach for single state systems undergoing phase separation and here we endow particles with multiple internal states governed by transition rates between each other:
\begin{equation} \label{eq: FP full}
    \begin{split}
    	\partial_t P(\vec{x},\vec{\sigma},t) =& \sum_{i=1}^{N} \Big[ D~\nabla_i\big[\nabla_i P(\vec{x},\vec{\sigma},t)  \\&+ \beta P(\vec{x},\vec{\sigma},t) \nabla_i  E(\vec{x},\vec{\sigma})\big] \\& + \sum_{\lbrace{\sigma_i'}\rbrace}^K
    	\Big(k_{\sigma_i' \sigma_i}(\vec{x})P(\vec{x},\vec{\sigma}_{/i},\sigma_i',t) \\& - k_{\sigma_i \sigma_i'}(\vec{x})P(\vec{x},\vec{\sigma}_{/i},\sigma_i,t) \Big) \Big]
     \end{split}
\end{equation}
where $\vec{\sigma}_{/i} = (\sigma_1, ...,\sigma_{i-1},\sigma_{i+1}, ...,\sigma_N)$ indicates the set of all chemical states apart from $\sigma_i$, and $\beta = 1/(k_B T)$, with $k_B$ the Boltzmann constant.

\begin{figure*}
\centering
    \includegraphics[width=.25\paperwidth]{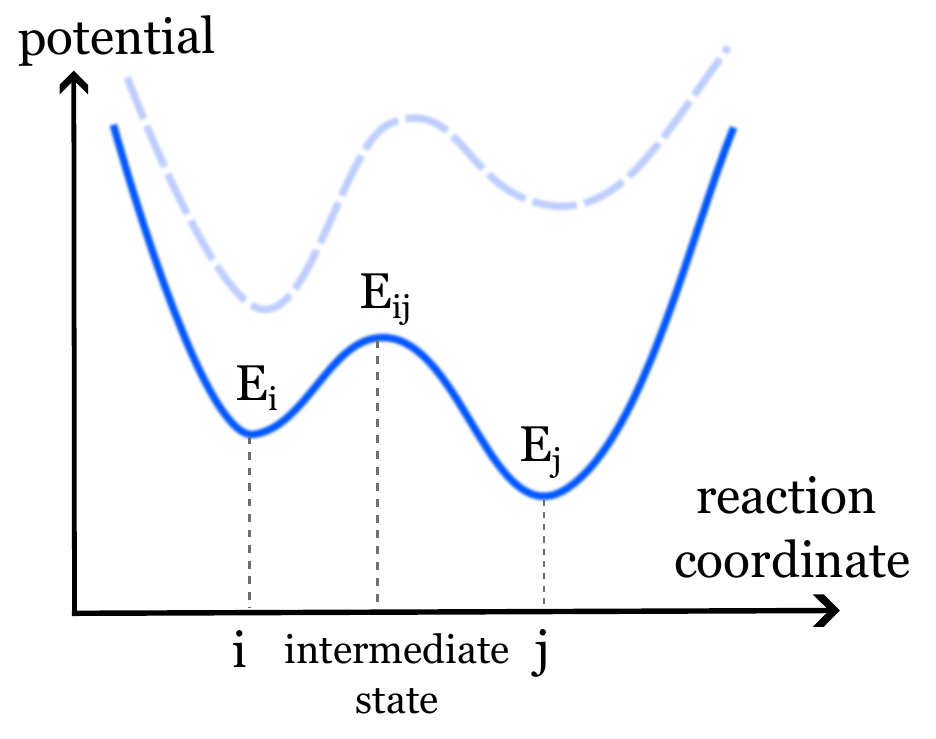}
    \hfill
    \includegraphics[width=.2\paperwidth]{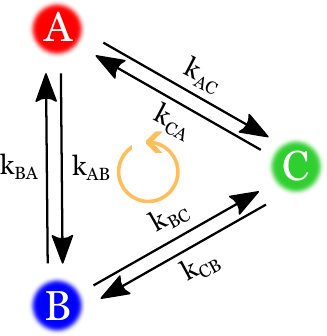}
    \hfill
    \includegraphics[width=.25\paperwidth]{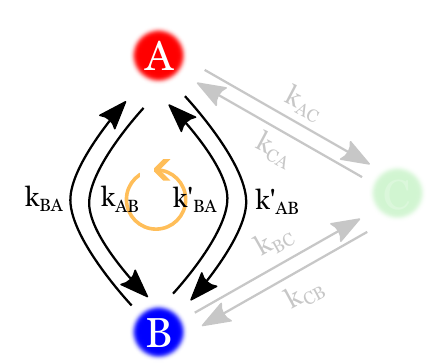}
\caption{The model. Left: Potential landscape of a chemical reaction changing state i to state j and \textit{vice versa}. Dashed line shows the potential landscape without the influence of the background potential.  Middle: Active chemical cycle with three chemical species. Right: Coarse-grained chemical cycle where one of the species is getting replaced by a second pathway with different reaction rates than on the first one.}
\label{figmodel}
\end{figure*}

Under the assumption that all reactions can be seen as activation processes over energy barriers,
the rate of the chemical transition of particle $i$ from state $\sigma_i$ to state $\sigma_i'$ via a transition state $[\sigma_i\sigma'_i]$ is
\begin{equation} \label{eq: overkill chem rates}
k_{\sigma_i \sigma_i'}(\vec{x}) = k^0_{\sigma_i \sigma_i'} e^{-\beta [E(\vec{x},\vec{\sigma}_{/i},[\sigma_i\sigma_i'])- E(\vec{x},\vec{\sigma}_{/i},\sigma_i)] }e^{\beta \mu^f_{\sigma_i \rightarrow \sigma_i'}} \quad ,
\end{equation}
where, for clarity, we separated the state of the $i$th particle from all the others.
This rate is composed of three factors. The first one, $k^0_{\sigma_i \sigma_i'}$, is the intrinsic rate of the transition as it would happen if the particle were not affected by external interactions, and it depends on the internal energies of each state and of the barrier. Yet, in the presence of external potentials and interactions, as detailed in Eq.~\ref{energies}, both the energies of the states and of the barriers change (Fig.~\ref{figmodel}A), resulting in the modifications captured by the second factor in (\ref{eq: overkill chem rates}). Here, the rates take the customary Kramers form with an energy barrier between $\sigma_i$ and $\sigma_i'$, two local minima of the modified energy landscape (see App.~\ref{sec: app FP to free energy} for details). Finally, the third factor describes the further modification of the rate in the presence of an active process that affects the reaction in selected directions, associated to an effective chemical potential $\mu^f_{\sigma_i \rightarrow \sigma_i'}$ (where $f$ stands for fuel).\\

In general, we are interested in the evolution of the concentrations of particles, which are just proportional to their probabilities obtained by marginalizing the $N$-body Fokker-Planck equation (\ref{eq: FP full}) over the coordinates and states of all the particles that are not of interest: 
\begin{equation}
\begin{split}
    &P(\vec{x}_i,\sigma_i,t) = 
   \sum_{ \lbrace \sigma_j=1,...K \rbrace_{j \ne i}} \int \prod_{j\neq i} d\vec{x}_j P(\vec{x},\vec{\sigma},t) 
\end{split}
\end{equation}

In the following, for brevity we define the notation 
\begin{equation}
    \SumInt_{/i} \equiv  \sum_{ \lbrace \sigma_j=1,...K \rbrace_{j \ne i}} \int \prod_{j\neq i} d\vec{x}_j \quad .
\end{equation}
and we drop the vector sign for single-particle positions. The consequent marginalization of Eq.~\eqref{eq: FP full} leads to the following equation:
\begin{equation} \label{eq: marge}
        \begin{split}
    	    \partial_t &P(x_i, \sigma_i ,t) = D~\nabla_i\Big[\nabla_i P(x_i,\sigma_i,t) \\
            &+\beta P(x_i,\sigma_i,t) \SumInt_{/i}  \Big [P(\vec{x}_{/i},\vec{\sigma}_{/i}|{x_i,\sigma_i}) \nabla_i E(\vec{x}_{/i},x_i,\vec{\sigma}_{/i},\sigma_i)\Big] \\
            &+ \sum_{\sigma_i' \neq \sigma_i}^K \Big[-k^{eff}_{\sigma_i \sigma_i'}(x_i)P(x_i,\sigma_i,t) + k^{eff}_{\sigma_i' \sigma_i}(x_i)P(x_i,\sigma_i',t)\Big]
    	\end{split}
	\end{equation}
where we have defined 
\begin{equation} \label{eq: k eff}
	\begin{split}
	    k^{eff}_{\sigma_i \sigma_i'}(x_i)= &
    	    k_{\sigma_i \sigma_i'}^0 \SumInt_{/i}  \Big(P(\vec{x}_{/i},\vec{\sigma}_{/i}|{x_i,\sigma_i}) \\&
    	    e^{-\beta E(\vec{x}_{/i},x_i,\vec{\sigma}_{/i},[\sigma_i \sigma_i']) + \beta E(\vec{x}_{/i},x_i,\vec{\sigma}_{/i},\sigma_i)}\Big)
	\end{split}
\end{equation}
and we have used $P(\vec{x},\vec{\sigma}) = P(\vec{x}_{/i},\vec{\sigma}_{/i}|{x_i,\sigma_i}) P(x_i,\sigma_i)$.

The presence of the conditional probability $P(\vec{x}_{/i},\vec{\sigma}_{/i}|{x_i,\sigma_i})$ maintains the same complexity as the full $N$-particle Fokker-Planck equation, and we must thus use approximations to reduce it to a more manageable form. While it is for sure not the only possibility, here we are postulating that the $N-1$ particles that we are tracing over adapt to the evolved state of particle $i$ with a much faster rate, eventually settling in their equilibrium state given $x_i$ and $\sigma_i$.\\
As a consequence (see App.~\ref{sec: app FP to free energy}),
\begin{equation}
P(\vec{x}_{/i},\vec{\sigma}_{/i}|{x_i,\sigma_i}) = \frac{e^{-\beta E(\vec{x}_{/i},x_i,\vec{\sigma}_{/i},\sigma_i)}}{Z(x_i,\sigma_i)}
\end{equation}
where
\begin{equation}
Z(x_i,\sigma_i) = \SumInt_{/i} {e^{-\beta E(\vec{x}_{/i},x_i,\vec{\sigma}_{/i},\sigma_i)}}.
\end{equation}
From this partition function of $N-1$ particles given the state of particle $i$, we can define a corresponding \textit{conditional free energy}:
\begin{equation}
F_{\text{cond}}(x_i,\sigma_i) = - k_B T \ln Z(x_i,\sigma_i).
\end{equation}

We similarly define the \textit{barrier partition function} 
\begin{eqnarray}
Z(x_i,[\sigma_i\sigma_i']) &=& \SumInt_{/i} {e^{-\beta E(\vec{x}_{/i},x_i,\vec{\sigma}_{/i},[\sigma_i\sigma_i'])}} \nonumber \\ &=& e^{-\beta F_{\text{cond}}(x_i,[\sigma_i\sigma_i'])} \quad .
\end{eqnarray}

Using these expressions, after some trivial algebra the effective transition rate Eq.~\eqref{eq: k eff} becomes
\begin{equation} \label{eq: k eff F}
k^{eff}_{\sigma_i,\sigma_i'}(x_i) = k^0_{\sigma_i,\sigma_i'} e^{-\beta \left[ F_{\text{cond}}(x_i,[\sigma_i\sigma_i']) - F_{\text{cond}}(x_i,\sigma_i)\right]}
\end{equation}
Taking into account that all particles in the same state $\sigma$ have the same probability distribution, and that their concentration is $c(x,\sigma) = N P(x,\sigma)$, Eq.~\eqref{eq: marge} eventually reduces to 
\begin{equation} \label{eq: 1-body FP}
\begin{split}
    \partial_t &c(x, \sigma ,t) = D~\nabla\Big[\nabla c(x,\sigma,t) \\&+\beta c(x,\sigma,t) \nabla F_{\text{cond}}(x,\sigma) )\Big] \\&
    + \sum_{\sigma' \neq \sigma}^K \Big[-k^{eff}_{\sigma \sigma'}(x)c(x,\sigma,t) + k^{eff}_{\sigma' \sigma}(x)c(x,\sigma',t)\Big]
\end{split}
\end{equation}
as described in more detail in Appendix A. The diffusive part of Eq.\ref{eq: 1-body FP} is the usual generalized Cahn-Hilliard equation, where $F_{\text{cond}}$ is the chemical potential, since it is the derivative of the free energy with respect to the concentration. 

Note that the Kramers form of the rates with free energies instead of energies emerges as a consequence of choosing Kramers rates for the microscopic reactions in the full, $N$-body description of the system. In this respect, it is worth highlighting that while the ratio between the forward and backward effective reaction rates in Eq.~\eqref{eq: k eff F} expectedly corresponds to the exponential of the free-energy difference, the form of the rates themselves cannot be chosen arbitrarily with the sole constraint of obeying detailed balance. Instead, a careful treatment of the microscopic reaction is in principle necessary to know which functional form the rates take. In particular, we emphasise the relevance of the barrier, as it will be shown in the remainder of the work to play a crucial role. Furthermore, it is also worth noting that the same free energy, $F_{\text{cond}}(x_i,\sigma_i)$, shaped by the local environment,  dictates both the chemical rates and the drift force, underscoring the intimate connection between the different parts of 
Eq.~\ref{eq: 1-body FP}.\\

In this work, we are interested in the relation between spatial inhomogeneities, non-equilibrium conditions and the presence of spatially extended fluxes at steady state ($\partial_t c(x,\sigma) = 0$). In what follows, without loss of generality, we will describe one-dimensional systems in a finite box of size $2L$. 

%%%%%%%%%%%%%%%%%%%%%%
\section{Necessary conditions for spatial fluxes} \label{sec: obtain fluxes}

While the full steady-state solution of the system must satisfy Eq.~\eqref{eq: 1-body FP} with $\partial_t c(x,\sigma) = 0$, our strategy for this Section is to solve first the chemical steady state locally, and then check whether the ensuing solution also separately satisfies the purely diffusion-drift part. If that is the case, then there are no spatially extended fluxes, otherwise they are present by necessity.

\textit{Chemical reaction network without cycles}: Let us first consider a system in which particles can be in two states, $\sigma=\{A, B\}$, such that the conversion of A into B is fuelled by a chemical potential $\mu_{A\rightarrow B}$. The local purely chemical steady state is
\begin{equation}\label{eq: ratio 2 species}
\begin{split}
    c_A(x)&=c_B(x)\frac{k_{BA}(x)}{k_{AB}(x)}\\&= c_B(x)\frac{k^0_{BA}}{k^0_{AB}}e^{-\beta(F_{A}(x)-F_{B}(x))}e^{-\beta \mu_{A\rightarrow B}} \quad.
    \end{split}
\end{equation}
where we have dropped "$\text{cond}$" from the free energy and written the chemical state in the subscript for the sake of simplicity. It is important to stress that, in the absence of cycles, there can be no fluxes on the network at steady-state, and the solution must respect the mathematical detailed balance, even if there is an underlying energy-consuming process ($\mu_{A \to B}$) driving some of the transitions.

Substituting this relation in the steady-state Eq.~\ref{eq: 1-body FP} for $A$, the chemical part vanishes by construction, and the spatial part is solved by the same spatial no-flux solution that solves the equation for B (see Appendix~\ref{sec: app no flux 2 species}), ultimately resulting in
\begin{eqnarray}
c_B(x) &=& c_0 e^{-\beta F_B(x)} \nonumber \\
c_A(x) &=& c_0 e^{-\beta F_A(x)} e^{-\beta \mu_{A\rightarrow B}}
\end{eqnarray}
where $c_0$ is fixed by the overall normalisation.  The presence of a non-equilibrium driving that does not depend on space is thus not sufficient to generate fluxes because it simply changes some \textit{apparent} energies of the system by a constant multiplicative term. 

Similar arguments can be used in the case of more complex chemical reaction networks, as long as they lack cycles: the concentrations of the states on the leaves of the network can be expressed in terms of the concentrations of the nodes they are connected to, in analogy to (\ref{eq: ratio 2 species}), restoring to the effective local detailed balance condition used above The iteration of this procedure reduces all equations to the equation of the root of the tree-like reaction network (see Appendix~\ref{app: tree}), which can then be solved for spatial no-flux conditions.
We have thus highlighted the first requirement that is necessary for the presence of spatially extended fluxes: the chemical reaction network must have cycles.
Yet, the presence of cycles is not sufficient in itself for the presence of fluxes.

\textit{General chemical reaction networks}: For a general chemical reaction network, the local chemical steady-state for particles in state $i$ can be written using the spanning-tree form of the solution \citep{schnakenberg}:
\begin{equation}
c_\sigma(x) = \frac{\sum_{\lbrace \mathcal{T} \rbrace} {\prod_{l \in \mathcal{T}}} k_{l}^{\sigma}(x)}{\mathcal{N}(x)}
\end{equation}

where $\lbrace \mathcal{T} \rbrace$ is the set of spanning trees over the network, $k_l^{\sigma}$ is the rate of the reaction on the network edge $l$ belonging to $\mathcal{T}$, directed toward $\sigma$, and $\mathcal{N}(x)$ is a general normalisation that does not depend on the state $\sigma$.
Using (\ref{eq: k eff F}), after some algebra it is possible to write this expression relative to a reference state concentration (say, the concentration of state $1$) as
\begin{eqnarray}\label{eq: tree eq}
c_\sigma(x) &=& c_1(x) e^{-\beta\left(F_\sigma(x)-F_1(x)\right)} \frac{\sum_{\lbrace \mathcal{T} \rbrace} e^{-\beta \mu_{1\sigma}} {\prod_{l \in \mathcal{T}}} k_{l}^{1}(x)} 
{\sum_{\lbrace \mathcal{T} \rbrace} \prod_{l \in \mathcal{T}} k_{l}^{1}(x)} \nonumber \\
&=& c_1(x) e^{-\beta\left(F_\sigma(x)-F_1(x)\right)} S_{1\sigma}(x)
\end{eqnarray}
where $\mu_{1\sigma}$ is the total %fuel 
chemical potential due to the fuel accumulated over the chemical reaction pathway from states $1$ and $\sigma$ (for the detailed derivation see Appendix~\ref{app: cycle no-flux}).\\

According to Eq.~\eqref{eq: k eff F}, the product of the rates over the edges of each spanning tree depends on the exponential of the sum of all (free) energies, which is the same for each tree and thus cancels in Eq.~\eqref{eq: tree eq}. Consequently $S_{1\sigma}(x)$ depends only on the fuel chemical potentials associated to the reactions, and on the barriers of the transitions. Substituting this expression in the steady-state equation for $c_i(x)$, the chemical reactions part vanishes by construction, while the spatial part acts solely on the reference concentration, i.e., $c_1(x)$ in this case. The resulting equation coincides with the stationary condition of $c_1(x)$, and thus can be consistently solved, only if $\nabla {S_{1\sigma}(x)}=0$, that is, $S_{1\sigma}(x)$ is constant in space (Appendix~\ref{app: cycle no-flux}). This condition is satisfied in three cases: \textit{i)} the system is at equilibrium ($\mu_{1\sigma}=0,   \forall \sigma $) and thus $S_{1\sigma}(x)=1$, or \textit{ii)} the barriers do not depend on space, or \textit{iii)} the barrier energies  all depend on space in the same way, thus leading to no spatial dependence when computing the rate (see Eq.~\eqref{eq: tree eq}).
The two-state system and the cycle-less cases are trivially recovered because there is only one spanning tree and, as such, the system satisfies an effective equilibrium, as detailed above.

We can thus summarise the necessary requirements for diffusive fluxes:
\begin{itemize}
\item \textbf{Activity}: At least one of the chemical reactions has to be out of equilibrium.
\item \textbf{Chemical cycles}: Without a chemical cycle, the systems adjusts into a pseudo-equilibrium state. 
\end{itemize}
The presence of activity and chemical cycles ensures that there will be chemical fluxes, \textit{i.e.}, the chemical steady state does not locally satisfy detailed balance. However, these fluxes only govern the local conversion of species and do not necessarily turn into spatial fluxes, unless a third condition is satisfied:
\begin{itemize}
\item \textbf{Spatial dependency of energies}. The energy landscape of the chemical reaction, and in particular the energy barriers of the transitions, must depend on space.
\end{itemize}

The minimal system with a cycle contains three species, $A$, $B$ and $C$. This cycle can describe, for example, the conversion of $A$ into $B$ via an active and a passive pathway, so that the third species $C$ represents a composite state of $A$ interacting with an enzyme. This three-state system can be studied either as just sketched or coarse-grained to a two-state system, with a passive pathway and an active pathway that is absorbing all interaction with the enzyme and the $C$ state (see Appendix \ref{sec: app coarsegraining} and Fig.~\ref{figmodel}). It is important to note that, in the presence of multiple reaction pathways, a two-state system shows cycles, and can thus exhibit fluxes. 

\section{Spatial structure of the fluxes} \label{sec: max fluxes}

Having established our formalism, we can now address the problem of the structure of fluxes in the presence of a spatial partition of the system induced by non-homogeneous free energies $F(x,\sigma)$ and $F(x,[\sigma'\sigma])$. Here, we assume that the functional space-dependent form of the free energies is given as an external potential: $F(x,\sigma) = V(x,\sigma)$ and $F(x,[\sigma'\sigma]) = V(x,[\sigma'\sigma])$. From a biological perspective, such a scenario might capture, for example, the physics of a chemical reaction system where the different molecules (\textit{e.g.} substrates, enzymes and products) are sensitive to the presence of phase-separated (condensate) regions that can selectively attract or repel them. Alternatively, the active maintenance of liquid-liquid phase separation might be captured by our framework, with the \textit{caveat} that the inhomogeneous free energies would depend on some of the concentrations themselves, thus leading in principle to self-consistent equations. At steady-state, and within a certain degree of approximation, we are confident that our conclusions are valid also in this setting.  

In what follows, we study a three-state system with the states $\sigma = \{A, B, C\}$. To illustrate our results, we restrict our analysis to the specific case when only one reaction, from A to C, is actively driven.

\subsection{Step potentials}

As the simplest model for an interface we use a step potential of the form 
\begin{equation}
V_\sigma(x) = s_\sigma V_\sigma^0 \Theta(x)
\end{equation}
where $V_\sigma^0$ determines the amplitude of the potential, $s_\sigma$ determines its sign and $\Theta(x)$ is the Heaviside Theta function. 
We write the energy of the barrier as 
\begin{equation}
V_{[\sigma\sigma']}(x) = V_{[\sigma\sigma']}^0(C_{[\sigma\sigma']} + \Theta(x))
\end{equation}
with the offset $C_{[\sigma\sigma']}$ that guarantees that the energy barriers are always higher than the initial and final energies in the reactions, a condition necessary for the Kramers approximation to be legitimate.\\ 
The analytical solutions for Eq.~\eqref{eq: 1-body FP} are obtained by solving it for $x<0$ and $x>0$ separately and connecting the solutions at $x = 0$ (for more details, see App.~\ref{app: BCs}). 
The solution on the left of the interface reads:
\begin{equation} \label{eq: concentration step}
    \vec{c}^{L}(x) = A^{L}_0 \vec{v}^{L}_0 + \sum_{i = 1,2} A^{L}_i \cosh{\Big(\sqrt{\lambda^{L}_i} (x + L)\Big)} \vec{v}^{L}_i 
\end{equation}
and similarly on the right of the interface, where the argument of the hyperbolic cosine changes to $x-L$ and $A^{L}_i \to A^{R}_i$ are constants that are fixed by the boundary conditions at the interface. The parameters $\lambda_i^{L/R}$ for $i=1,2$ are the two non-zero eigenvalues of the chemical transition matrix on each side of the interface (the same approach can be used to find spherically symmetric solutions in three dimensions, which is more pertinent to real condensates, see App.~\ref{sec: app spherical coords}). 

\begin{figure*}[p!]
\centering
    \includegraphics[width=.8\paperwidth]{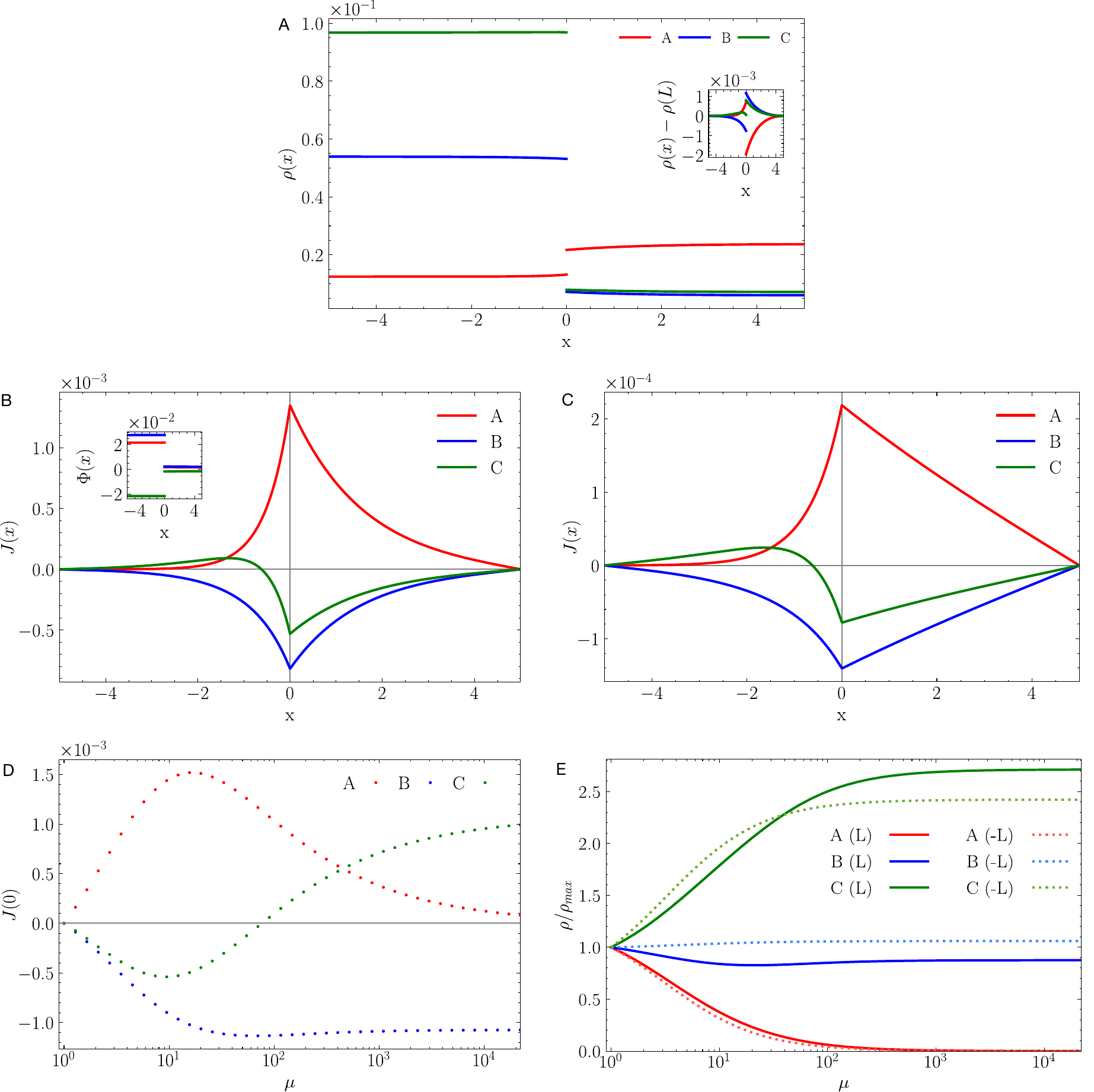}
\caption{A: Concentration profiles of the three species for a step potential. Unless mentioned otherwise, we set the parameters for all calculations to be $E_A = -0.5$, $ E_B = 2.$, $E_C = 2.5$, $E_{AB} = 2.$, $E_{BC} = 2.5$, $E_{AC}= 3.$, $ s_B=s_C = -s_A = 1$, $\mu_{A\rightarrow C} = 2$ and $C_{[AB]} = C_{[BC]} = C_{[AC]} =0.5$. Inset: Profiles of `excess' concentration, calculated as the difference of the concentration value at x and the concentration value at the respective boundary of the system. B: Spatial fluxes of the three species. Inset: Chemical fluxes of the three species. C: Fluxes for $E_{AB} = 4.$, $E_{BC} = 4.5$, $E_{AC}= 5.$, resulting in a linear profile for $x>0$. D: Diffusive fluxes at $x = 0$ as a function of chemical potential $\mu_{A\rightarrow C} \equiv \mu$. E: Concentrations at $x = L$ (solid) and $x = -L$ (dotted) as a function of $\mu_{A\rightarrow C}$, normalized to the respective equilibrium concentration values ($\mu = 0$). As seen in A, $\rho(x)$ varies little on each side of the step, so it suffices to study its behaviour at the extremities of the box.} 
\label{Fig:stepfct1}
\end{figure*}

Representative concentration profiles are shown in Fig.~\ref{Fig:stepfct1}A (the values of the used parameters are reported in the figure caption). Because of the structure of the solution (\ref{eq: concentration step}), the largest deviations from a constant solution are found at the interface (inset in Fig.~\ref{Fig:stepfct1}A), and as a consequence the spatial fluxes, that for the step potential are just the first derivatives of the concentrations, are maximal at the interface and decay away from it (Fig.~\ref{Fig:stepfct1}B). It is important to stress here that the absence of spatial fluxes far away from the interface does not imply the absence of fluxes over the chemical reaction network (inset in Fig.~\ref{Fig:stepfct1}B). Rather, away from the interface reactions are chemically compensated and there is no "spill-over" generating fluxes in space.

Since the potentials are constant on both sides of the interface, there is no drift term, and as a consequence the fluxes are purely diffusive and decay in space according to the reaction-diffusion length $l^{rd}_{L/R}$, which is asymptotically dominated by the smallest (non-zero) eigenvalue of the rate matrix of the chemical reactions,  $l^{rd}_{L/R} = (\lambda_{min}^{L/R} )^{-1/2}$ (for simplicity, here we consider all diffusion constants to be identical, and the reaction-diffusion length is actually related to the diffusion-normalized chemical rates matrix; the same treatment can be generalized to systems with state-dependent diffusion constants.
In regions where chemical rates are sufficiently slow compared to the diffusive rate, the diffusive fluxes show an almost linear behaviour from the system boundaries ($x=\pm L$) to the interface $x=0$ (Fig.~\ref{Fig:stepfct1}C). This result stems from Eq.~\eqref{eq: concentration step}: the fluxes are proportional to the derivative of the concentrations, so they will be proportional to $\sinh((x \pm L)/l_{L/R}^{rd})$, which is roughly linear over the full spatial range if $L/l_{L/R}^{rd} \ll 1$. In such a situation, fluxes can ``penetrate'' fully in at least one of the two regions.

The behavior of the fluxes with respect to the chemical driving is shown in Fig.~\ref{Fig:stepfct1}D, where we report their value at the interface. Fluxes vanish at equilibrium ($\mu_{A\rightarrow C} = 0$), as it must be, and increase (in absolute value) proportionally to $\mu_{A\rightarrow C}$. While this relation is expected for small values of $\mu_{A\to C}$ as from a first-order perturbative approach, it breaks down for higher values of $\mu_{A\rightarrow C}$, becoming non-monotonic: one of the fluxes ($J_A$) vanishes for large non-equilibrium driving, and one ($J_C$) changes signs, namely, it changes direction. Concomitantly, one of the chemical species approaches extinction (Fig.~\ref{Fig:stepfct1}E). While this behavior can be understood because as $\mu_{A \to C}$ grows state A is depleted and thus it cannot give rise to significant spatial fluxes, it highlights how, in more general scenarios and for more complex chemical reaction networks, the behavior for intermediate and large driving can be highly non-trivial.

Since in most of the literature, spatial inhomogeneities and chemical reactions are studied for two-state systems. In Appendix~\ref{sec: app coarsegraining} we show how to coarse-grain a three-state chemical reaction network with cycles to a two-state ones, with cycles due to the presence of two different reaction pathways between the two remaining states, and we further re-derive the results of this section for that setup.

\subsection{Sigmoidal potentials}
To study the effects of a finite interface width, and to create a more realistic model of a biological interface, we turn to a sigmoidal potential $V(x,\sigma)$ with the functional form here of a hyperbolic tangent. This choice, that introduces a smooth interface of controlled width, is inspired by the profile of the interface of phase-separated regions \citep{ActiveEmulsions}, which would provide the relevant background energy for the molecules involved in the chemical reaction network. \\

We thus choose potentials of the form
\begin{equation} \label{eq: tanh potential}
V_\sigma(x) = 0.5~V^0_\sigma (1 + s_\sigma \tanh(n \cdot x)) 
\end{equation}
where $V^0_\sigma$ is controlling the potential amplitude and $n$ is its slope. The sign of the hyperbolic tangent is set by $s_\sigma \in \{-1,1\}$. 
For the energy barriers, we choose 
\begin{equation} \label{eq: tanh barrier}
V_{[\sigma'\sigma]}(x) = 0.5~V^0_{[\sigma'\sigma]} (C_{[\sigma'\sigma]} + \tanh(n \cdot x))
\end{equation}
where $C_{[\sigma'\sigma]}$ is an additional offset which guarantees that the energy barriers are always higher than the initial and final energies in the reactions, a condition necessary for the Kramers approximation to be legitimate.

\begin{figure*}
\centering
    \includegraphics[width=.8\paperwidth]{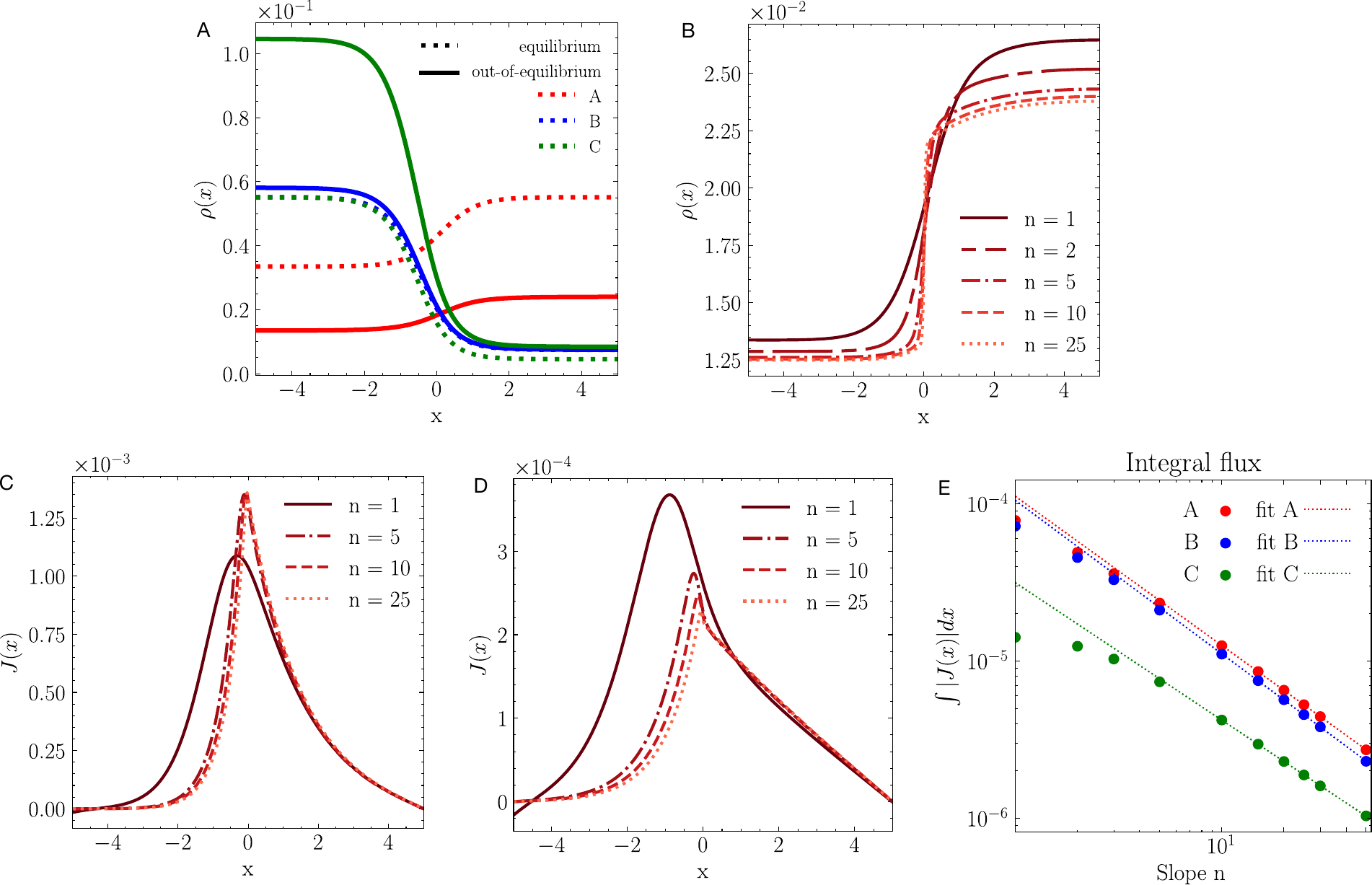}
    
\caption{Sigmoidal potentials for the same parameters as the step potentials, with $C_{[AB]} = C_{[BC]} = C_{[AC]} = 3$ (a change to ensure the barrier potentials have the same values at the system boundaries). A: Comparison of the concentration profiles out of equilibrium with the equilibrium case (dashed lines) for the three species A, B and C for $n_A = n_B = n_C = n = 1$. B: Concentration profiles of species A with different tanh slopes $n$. C: Flux profiles of species A with different tanh slopes $n$. D: Concentration profiles of species A for $E_{AB} = 4.$, $E_{BC} = 4.5$, $E_{AC}= 5.$, resulting in an approximately linear profile for $x>0$. E: Integral of the absolute value of the flux excess over the flux found for a step potential, averaged over the system length, as a function of slope $n$. The dashed line shows the fit to the last 6 points of the function $y(n) = n^a 10^b$. The retrieved fitting parameters a are very close to -1.}
\label{Fig:TanhRhos}
\end{figure*}

In the absence of analytical solutions of Eq.~\eqref{eq: 1-body FP} with these potentials, we solve it numerically (see Appendix \ref{sec: app num sol} for a description of the method).
The resulting concentrations have, as expected, a sigmoidal shape, centred at $x=0$ (Fig.~\ref{Fig:TanhRhos}A), different from the equilibrium ones ($\mu_{AC}=0$, Fig.~\ref{Fig:TanhRhos}A dashed lines). The behavior of the concentrations as a function of the sharpness of the interface is shown in Fig.~\ref{Fig:TanhRhos}B (only the results for the concentration of species A are shown) and expectedly, they approach a limit profile for $n \to \infty$, corresponding to step potentials (here we have chosen the same $n$ for all species, while of course it could be different for each one of them). As the interface becomes sharper the fluxes also approach the step-function limit behaviour. 
 
We again observe linear fluxes for the same parameter regime of energies and energy barriers as for the step function (Fig.~\ref{Fig:TanhRhos}D). This parameter range corresponds to  reaction-diffusion lengths much bigger than the system size, highlighting the consistency with the solution with the step potential. 

In order to assess whether a smoother interface is increasing or decreasing the net flux with respect to the step potential case, we study the excess average total flux magnitude. It is calculated as an integral of the absolute flux value over the system length for a slope $n$ minus the same quantity for $n \rightarrow \infty$ (step function). We find that it is positive and proportional to $n^{-t}$ with exponents $t$ very close to 1 (for Fig.~\ref{Fig:TanhRhos}E, $t_A = 0.95$, $t_B = 0.98$, $t_C = 0.87$). This shows that a smoother interface corresponds to higher spatial fluxes, a result that can possibly be assessed in experimental observations of known active condensates.

\section{Discussion} \label{sec: discussion} 
For condensates to act as chemical factories, there must be an influx of substrates and outflux of products, implying that fluxes are necessarily present across their interfaces. Likewise, when structural components of condensates are actively modified, so to change their state and condensation propensity, outgoing and ingoing fluxes of molecules in different states have to be expected.

Here we focused on a spatial description of the condensates to explore the conditions that are necessary for the presence of such fluxes, and their spatial structure. We have found that two non-trivial conditions for the presence of fluxes  are the presence of cycles in the  chemical reaction networks, and the space and state dependence of the energy barriers governing the transition rates. Of course, and more trivially, when these conditions are met, fluxes are present only if reaction are driven away from thermodynamic equilibrium by energy consumption (some $\mu_{\sigma \sigma'} \neq 0$).
 
If this set of conditions is satisfied, then fluxes are present, and are mostly localized at the condensate interface, extending on both sides by, essentially, the reaction-diffusion length $l^{rd}$. As a consequence, regions of the system that are far away (relative to $l^{rd}$) from the interface do not contribute to the transport of substrates and products, because the non-equilibrium fluxes over the chemical reaction network are fully chemically compensated, and do not leak in space.
Therefore, since only a limited region close to the interface can act as a chemical factory, it is tempting to speculate that maintaining the size of condensates limited, instead of letting them all coalesce as they would if they where just following the laws of equilibrium thermodynamics, might be useful to increase the overall condensate surface and the associated effective reactive region.
Furthermore, our results show that the net flux is decreasing with increasing interface slope, allowing to speculate that condensates functioning as chemical factories will be characterized by a smooth, fuzzy interface and negligible surface tension. 
 
In this respect, renewed attention should be devoted to studying the properties of condensates close to their interfaces in more detail to better characterize their molecular arrangements. Indeed, numerous works, both numerical \citep{Cho_2023, Farag} and experimental \citep{ishov2020coordination, wang2018intronless, Boeddeker, Folkmann_2021, wu2023single}, have shown that the surfaces of condensates have distinct physical properties from the bulk and the surroundings of the condensate. Furthermore, there is an increasing number of observations that aggregation and fibril formation rates are enhanced at the condensate interface \citep[and others]{Lipinski, ShenFUS}. Simulations at different levels of detail provide the same results (\cite{KnowlesAggregatesSurface, Farag}).

From a modelling perspective, our work shows the relevance of accounting for the spatial dependence of the chemical rates, as well as the explicit treatment of the energy barriers. Some authors (\textit{e.g.}, \cite{Kirschbaum_2021, laha}) choose $k^0_{ij}$ and $\mu_{i \rightarrow j}$ to be space-dependent, absorbing $\epsilon_{ij}(x)$ and $V_i(x)$, respectively, into their definitions. Comparing to such works, it should be kept in mind how the constraints we derived here are translated with differing definitions of the variables.
Furthermore, it is important to stress that there is an intimate connection between the spatial and chemical parts of the governing equation of the system, Eq.\ref{eq: 1-body FP}: space-dependent chemical transition rates should be accompanied by corresponding drift terms.

While this work has been devoted to elucidating the physical rules that are necessary for the appearance of spatial fluxes, and we have started to highlight some of their properties, it ultimately calls for a better experimental characterization  both of the biochemical networks underlying active processes in phase-separated systems, and of the interactions that determine the phase separation itself and, as a consequence, the perturbation of the chemical rates.  The consequent implementation of the present findings in more detailed models, possibly closer to biochemical/biological instances, will provide avenues to understand the functional advantages of certain condensates, of their compositions and of their size distributions.\\

\section*{Acknowledgements}
We thank Alessandro Barducci and David Zwicker for fruitful discussions, as well as Francesco Piazza for a critical reading of the manuscript. This work was supported by the Swiss National Science Foundation under grant CRSII5\_193740.

\clearpage

\bibliography{main}

\begin{thebibliography}{27}%
\makeatletter
\providecommand \@ifxundefined [1]{%
 \@ifx{#1\undefined}
}%
\providecommand \@ifnum [1]{%
 \ifnum #1\expandafter \@firstoftwo
 \else \expandafter \@secondoftwo
 \fi
}%
\providecommand \@ifx [1]{%
 \ifx #1\expandafter \@firstoftwo
 \else \expandafter \@secondoftwo
 \fi
}%
\providecommand \natexlab [1]{#1}%
\providecommand \enquote  [1]{``#1''}%
\providecommand \bibnamefont  [1]{#1}%
\providecommand \bibfnamefont [1]{#1}%
\providecommand \citenamefont [1]{#1}%
\providecommand \href@noop [0]{\@secondoftwo}%
\providecommand \href [0]{\begingroup \@sanitize@url \@href}%
\providecommand \@href[1]{\@@startlink{#1}\@@href}%
\providecommand \@@href[1]{\endgroup#1\@@endlink}%
\providecommand \@sanitize@url [0]{\catcode `\\12\catcode `\$12\catcode `\&12\catcode `\#12\catcode `\^12\catcode `\_12\catcode `\%12\relax}%
\providecommand \@@startlink[1]{}%
\providecommand \@@endlink[0]{}%
\providecommand \url  [0]{\begingroup\@sanitize@url \@url }%
\providecommand \@url [1]{\endgroup\@href {#1}{\urlprefix }}%
\providecommand \urlprefix  [0]{URL }%
\providecommand \Eprint [0]{\href }%
\providecommand \doibase [0]{https://doi.org/}%
\providecommand \selectlanguage [0]{\@gobble}%
\providecommand \bibinfo  [0]{\@secondoftwo}%
\providecommand \bibfield  [0]{\@secondoftwo}%
\providecommand \translation [1]{[#1]}%
\providecommand \BibitemOpen [0]{}%
\providecommand \bibitemStop [0]{}%
\providecommand \bibitemNoStop [0]{.\EOS\space}%
\providecommand \EOS [0]{\spacefactor3000\relax}%
\providecommand \BibitemShut  [1]{\csname bibitem#1\endcsname}%
\let\auto@bib@innerbib\@empty
%</preamble>
\bibitem [{\citenamefont {Narayanan}\ \emph {et~al.}(2019)\citenamefont {Narayanan}, \citenamefont {Meriin}, \citenamefont {Andrews}, \citenamefont {Spille}, \citenamefont {Sherman},\ and\ \citenamefont {Cisse}}]{Cisse}%
  \BibitemOpen
  \bibfield  {author} {\bibinfo {author} {\bibfnamefont {A.}~\bibnamefont {Narayanan}}, \bibinfo {author} {\bibfnamefont {A.}~\bibnamefont {Meriin}}, \bibinfo {author} {\bibfnamefont {J.~O.}\ \bibnamefont {Andrews}}, \bibinfo {author} {\bibfnamefont {J.-H.}\ \bibnamefont {Spille}}, \bibinfo {author} {\bibfnamefont {M.~Y.}\ \bibnamefont {Sherman}},\ and\ \bibinfo {author} {\bibfnamefont {I.~I.}\ \bibnamefont {Cisse}},\ }\bibfield  {title} {\bibinfo {title} {A first order phase transition mechanism underlies protein aggregation in mammalian cells},\ }\href {https://doi.org/10.7554/eLife.39695} {\bibfield  {journal} {\bibinfo  {journal} {eLife}\ }\textbf {\bibinfo {volume} {8}},\ \bibinfo {pages} {e39695} (\bibinfo {year} {2019})}\BibitemShut {NoStop}%
\bibitem [{\citenamefont {Guilhas}\ \emph {et~al.}(2020)\citenamefont {Guilhas}, \citenamefont {Walter}, \citenamefont {Rech}, \citenamefont {David}, \citenamefont {Walliser}, \citenamefont {Palmeri}, \citenamefont {Mathieu-Demaziere}, \citenamefont {Parmeggiani}, \citenamefont {Bouet}, \citenamefont {{Le Gall}},\ and\ \citenamefont {Nollmann}}]{parABS}%
  \BibitemOpen
  \bibfield  {author} {\bibinfo {author} {\bibfnamefont {B.}~\bibnamefont {Guilhas}}, \bibinfo {author} {\bibfnamefont {J.-C.}\ \bibnamefont {Walter}}, \bibinfo {author} {\bibfnamefont {J.}~\bibnamefont {Rech}}, \bibinfo {author} {\bibfnamefont {G.}~\bibnamefont {David}}, \bibinfo {author} {\bibfnamefont {N.~O.}\ \bibnamefont {Walliser}}, \bibinfo {author} {\bibfnamefont {J.}~\bibnamefont {Palmeri}}, \bibinfo {author} {\bibfnamefont {C.}~\bibnamefont {Mathieu-Demaziere}}, \bibinfo {author} {\bibfnamefont {A.}~\bibnamefont {Parmeggiani}}, \bibinfo {author} {\bibfnamefont {J.-Y.}\ \bibnamefont {Bouet}}, \bibinfo {author} {\bibfnamefont {A.}~\bibnamefont {{Le Gall}}},\ and\ \bibinfo {author} {\bibfnamefont {M.}~\bibnamefont {Nollmann}},\ }\bibfield  {title} {\bibinfo {title} {Atp-driven separation of liquid phase condensates in bacteria},\ }\href {https://doi.org/https://doi.org/10.1016/j.molcel.2020.06.034} {\bibfield  {journal} {\bibinfo  {journal} {Molecular Cell}\ }\textbf {\bibinfo {volume} {79}},\ \bibinfo
  {pages} {293} (\bibinfo {year} {2020})}\BibitemShut {NoStop}%
\bibitem [{\citenamefont {Correll}\ \emph {et~al.}(2019)\citenamefont {Correll}, \citenamefont {Bartek},\ and\ \citenamefont {Dundr}}]{nucleolus}%
  \BibitemOpen
  \bibfield  {author} {\bibinfo {author} {\bibfnamefont {C.~C.}\ \bibnamefont {Correll}}, \bibinfo {author} {\bibfnamefont {J.}~\bibnamefont {Bartek}},\ and\ \bibinfo {author} {\bibfnamefont {M.}~\bibnamefont {Dundr}},\ }\bibfield  {title} {\bibinfo {title} {The nucleolus: A multiphase condensate balancing ribosome synthesis and translational capacity in health, aging and ribosomopathies},\ }\bibfield  {journal} {\bibinfo  {journal} {Cells}\ }\textbf {\bibinfo {volume} {8}},\ \href {https://doi.org/10.3390/cells8080869} {10.3390/cells8080869} (\bibinfo {year} {2019})\BibitemShut {NoStop}%
\bibitem [{\citenamefont {Wang}\ \emph {et~al.}(2018)\citenamefont {Wang}, \citenamefont {Wang}, \citenamefont {Wang}, \citenamefont {Chen}, \citenamefont {Shi},\ and\ \citenamefont {Cheng}}]{wang2018intronless}%
  \BibitemOpen
  \bibfield  {author} {\bibinfo {author} {\bibfnamefont {K.}~\bibnamefont {Wang}}, \bibinfo {author} {\bibfnamefont {L.}~\bibnamefont {Wang}}, \bibinfo {author} {\bibfnamefont {J.}~\bibnamefont {Wang}}, \bibinfo {author} {\bibfnamefont {S.}~\bibnamefont {Chen}}, \bibinfo {author} {\bibfnamefont {M.}~\bibnamefont {Shi}},\ and\ \bibinfo {author} {\bibfnamefont {H.}~\bibnamefont {Cheng}},\ }\bibfield  {title} {\bibinfo {title} {Intronless mrnas transit through nuclear speckles to gain export competence},\ }\href@noop {} {\bibfield  {journal} {\bibinfo  {journal} {Journal of Cell Biology}\ }\textbf {\bibinfo {volume} {217}},\ \bibinfo {pages} {3912} (\bibinfo {year} {2018})}\BibitemShut {NoStop}%
\bibitem [{\citenamefont {Ishov}\ \emph {et~al.}(2020)\citenamefont {Ishov}, \citenamefont {Gurumurthy},\ and\ \citenamefont {Bungert}}]{ishov2020coordination}%
  \BibitemOpen
  \bibfield  {author} {\bibinfo {author} {\bibfnamefont {A.~M.}\ \bibnamefont {Ishov}}, \bibinfo {author} {\bibfnamefont {A.}~\bibnamefont {Gurumurthy}},\ and\ \bibinfo {author} {\bibfnamefont {J.}~\bibnamefont {Bungert}},\ }\bibfield  {title} {\bibinfo {title} {Coordination of transcription, processing, and export of highly expressed rnas by distinct biomolecular condensates},\ }\href@noop {} {\bibfield  {journal} {\bibinfo  {journal} {Emerging Topics in Life Sciences}\ }\textbf {\bibinfo {volume} {4}},\ \bibinfo {pages} {281} (\bibinfo {year} {2020})}\BibitemShut {NoStop}%
\bibitem [{\citenamefont {Hondele}\ \emph {et~al.}(2020)\citenamefont {Hondele}, \citenamefont {Heinrich}, \citenamefont {De~Los~Rios},\ and\ \citenamefont {Weis}}]{hondele}%
  \BibitemOpen
  \bibfield  {author} {\bibinfo {author} {\bibfnamefont {M.}~\bibnamefont {Hondele}}, \bibinfo {author} {\bibfnamefont {S.}~\bibnamefont {Heinrich}}, \bibinfo {author} {\bibfnamefont {P.}~\bibnamefont {De~Los~Rios}},\ and\ \bibinfo {author} {\bibfnamefont {K.}~\bibnamefont {Weis}},\ }\bibfield  {title} {\bibinfo {title} {Membraneless organelles: phasing out of equilibrium},\ }\href@noop {} {\bibfield  {journal} {\bibinfo  {journal} {Emerging topics in life sciences}\ }\textbf {\bibinfo {volume} {4}},\ \bibinfo {pages} {343} (\bibinfo {year} {2020})}\BibitemShut {NoStop}%
\bibitem [{\citenamefont {Kedersha}\ \emph {et~al.}(2005)\citenamefont {Kedersha}, \citenamefont {Stoecklin}, \citenamefont {Ayodele}, \citenamefont {Yacono}, \citenamefont {Lykke-Andersen}, \citenamefont {Fritzler}, \citenamefont {Scheuner}, \citenamefont {Kaufman}, \citenamefont {Golan},\ and\ \citenamefont {Anderson}}]{Kedersha}%
  \BibitemOpen
  \bibfield  {author} {\bibinfo {author} {\bibfnamefont {N.}~\bibnamefont {Kedersha}}, \bibinfo {author} {\bibfnamefont {G.}~\bibnamefont {Stoecklin}}, \bibinfo {author} {\bibfnamefont {M.}~\bibnamefont {Ayodele}}, \bibinfo {author} {\bibfnamefont {P.}~\bibnamefont {Yacono}}, \bibinfo {author} {\bibfnamefont {J.}~\bibnamefont {Lykke-Andersen}}, \bibinfo {author} {\bibfnamefont {M.~J.}\ \bibnamefont {Fritzler}}, \bibinfo {author} {\bibfnamefont {D.}~\bibnamefont {Scheuner}}, \bibinfo {author} {\bibfnamefont {R.~J.}\ \bibnamefont {Kaufman}}, \bibinfo {author} {\bibfnamefont {D.~E.}\ \bibnamefont {Golan}},\ and\ \bibinfo {author} {\bibfnamefont {P.}~\bibnamefont {Anderson}},\ }\bibfield  {title} {\bibinfo {title} {Stress granules and processing bodies are dynamically linked sites of mrnp remodeling},\ }\href@noop {} {\bibfield  {journal} {\bibinfo  {journal} {The Journal of cell biology}\ }\textbf {\bibinfo {volume} {169}},\ \bibinfo {pages} {871} (\bibinfo {year} {2005})}\BibitemShut {NoStop}%
\bibitem [{\citenamefont {{Flory}}(1942)}]{Flory42}%
  \BibitemOpen
  \bibfield  {author} {\bibinfo {author} {\bibfnamefont {P.~J.}\ \bibnamefont {{Flory}}},\ }\bibfield  {title} {\bibinfo {title} {{Thermodynamics of High Polymer Solutions}},\ }\href {https://doi.org/10.1063/1.1723621} {\bibfield  {journal} {\bibinfo  {journal} {\jcp}\ }\textbf {\bibinfo {volume} {10}},\ \bibinfo {pages} {51} (\bibinfo {year} {1942})}\BibitemShut {NoStop}%
\bibitem [{\citenamefont {Huggins}(1942)}]{Huggins42}%
  \BibitemOpen
  \bibfield  {author} {\bibinfo {author} {\bibfnamefont {M.~L.}\ \bibnamefont {Huggins}},\ }\bibfield  {title} {\bibinfo {title} {Some properties of solutions of long-chain compounds.},\ }\href {https://doi.org/10.1021/j150415a018} {\bibfield  {journal} {\bibinfo  {journal} {The Journal of Physical Chemistry}\ }\textbf {\bibinfo {volume} {46}},\ \bibinfo {pages} {151} (\bibinfo {year} {1942})}\BibitemShut {NoStop}%
\bibitem [{\citenamefont {Weber}\ \emph {et~al.}(2019)\citenamefont {Weber}, \citenamefont {Zwicker}, \citenamefont {J{\"u}licher},\ and\ \citenamefont {Lee}}]{ActiveEmulsions}%
  \BibitemOpen
  \bibfield  {author} {\bibinfo {author} {\bibfnamefont {C.~A.}\ \bibnamefont {Weber}}, \bibinfo {author} {\bibfnamefont {D.}~\bibnamefont {Zwicker}}, \bibinfo {author} {\bibfnamefont {F.}~\bibnamefont {J{\"u}licher}},\ and\ \bibinfo {author} {\bibfnamefont {C.~F.}\ \bibnamefont {Lee}},\ }\bibfield  {title} {\bibinfo {title} {Physics of active emulsions},\ }\href@noop {} {\bibfield  {journal} {\bibinfo  {journal} {Reports on Progress in Physics}\ }\textbf {\bibinfo {volume} {82}},\ \bibinfo {pages} {064601} (\bibinfo {year} {2019})}\BibitemShut {NoStop}%
\bibitem [{\citenamefont {Kirschbaum}\ and\ \citenamefont {Zwicker}(2021)}]{Kirschbaum_2021}%
  \BibitemOpen
  \bibfield  {author} {\bibinfo {author} {\bibfnamefont {J.}~\bibnamefont {Kirschbaum}}\ and\ \bibinfo {author} {\bibfnamefont {D.}~\bibnamefont {Zwicker}},\ }\bibfield  {title} {\bibinfo {title} {Controlling biomolecular condensates via chemical reactions},\ }\href {https://doi.org/10.1098/rsif.2021.0255} {\bibfield  {journal} {\bibinfo  {journal} {Journal of The Royal Society Interface}\ }\textbf {\bibinfo {volume} {18}},\ \bibinfo {pages} {20210255} (\bibinfo {year} {2021})}\BibitemShut {NoStop}%
\bibitem [{\citenamefont {Bauermann}\ \emph {et~al.}(2022)\citenamefont {Bauermann}, \citenamefont {Laha}, \citenamefont {McCall}, \citenamefont {Jülicher},\ and\ \citenamefont {Weber}}]{Bauermann}%
  \BibitemOpen
  \bibfield  {author} {\bibinfo {author} {\bibfnamefont {J.}~\bibnamefont {Bauermann}}, \bibinfo {author} {\bibfnamefont {S.}~\bibnamefont {Laha}}, \bibinfo {author} {\bibfnamefont {P.~M.}\ \bibnamefont {McCall}}, \bibinfo {author} {\bibfnamefont {F.}~\bibnamefont {Jülicher}},\ and\ \bibinfo {author} {\bibfnamefont {C.~A.}\ \bibnamefont {Weber}},\ }\bibfield  {title} {\bibinfo {title} {Chemical kinetics and mass action in coexisting phases},\ }\href {https://doi.org/10.1021/jacs.2c06265} {\bibfield  {journal} {\bibinfo  {journal} {Journal of the American Chemical Society}\ }\textbf {\bibinfo {volume} {144}},\ \bibinfo {pages} {19294} (\bibinfo {year} {2022})},\ \bibinfo {note} {pMID: 36241174}\BibitemShut {NoStop}%
\bibitem [{\citenamefont {Osmanovic}\ and\ \citenamefont {Franco}(2022)}]{osmanovic2022chemical}%
  \BibitemOpen
  \bibfield  {author} {\bibinfo {author} {\bibfnamefont {D.}~\bibnamefont {Osmanovic}}\ and\ \bibinfo {author} {\bibfnamefont {E.}~\bibnamefont {Franco}},\ }\href@noop {} {\bibinfo {title} {Chemical reaction motifs driving non-equilibrium behaviors in phase separating materials}} (\bibinfo {year} {2022}),\ \Eprint {https://arxiv.org/abs/2207.10135} {arXiv:2207.10135 [cond-mat.soft]} \BibitemShut {NoStop}%
\bibitem [{\citenamefont {Zwicker}(2022)}]{Zwicker_2022}%
  \BibitemOpen
  \bibfield  {author} {\bibinfo {author} {\bibfnamefont {D.}~\bibnamefont {Zwicker}},\ }\bibfield  {title} {\bibinfo {title} {The intertwined physics of active chemical reactions and phase separation},\ }\href {https://doi.org/10.1016/j.cocis.2022.101606} {\bibfield  {journal} {\bibinfo  {journal} {Current Opinion in Colloid Interface Science}\ }\textbf {\bibinfo {volume} {61}},\ \bibinfo {pages} {101606} (\bibinfo {year} {2022})}\BibitemShut {NoStop}%
\bibitem [{\citenamefont {Archer}\ and\ \citenamefont {Evans}(2004)}]{Archer2004}%
  \BibitemOpen
  \bibfield  {author} {\bibinfo {author} {\bibfnamefont {A.~J.}\ \bibnamefont {Archer}}\ and\ \bibinfo {author} {\bibfnamefont {R.}~\bibnamefont {Evans}},\ }\bibfield  {title} {\bibinfo {title} {Dynamical density functional theory and its application to spinodal decomposition},\ }\href@noop {} {\bibfield  {journal} {\bibinfo  {journal} {The Journal of chemical physics}\ }\textbf {\bibinfo {volume} {121}},\ \bibinfo {pages} {4246} (\bibinfo {year} {2004})}\BibitemShut {NoStop}%
\bibitem [{\citenamefont {Schnakenberg}(1976)}]{schnakenberg}%
  \BibitemOpen
  \bibfield  {author} {\bibinfo {author} {\bibfnamefont {J.}~\bibnamefont {Schnakenberg}},\ }\bibfield  {title} {\bibinfo {title} {Network theory of microscopic and macroscopic behavior of master equation systems},\ }\href@noop {} {\bibfield  {journal} {\bibinfo  {journal} {Reviews of Modern physics}\ }\textbf {\bibinfo {volume} {48}},\ \bibinfo {pages} {571} (\bibinfo {year} {1976})}\BibitemShut {NoStop}%
\bibitem [{\citenamefont {Cho}\ and\ \citenamefont {Jacobs}(2023)}]{Cho_2023}%
  \BibitemOpen
  \bibfield  {author} {\bibinfo {author} {\bibfnamefont {Y.}~\bibnamefont {Cho}}\ and\ \bibinfo {author} {\bibfnamefont {W.~M.}\ \bibnamefont {Jacobs}},\ }\bibfield  {title} {\bibinfo {title} {Tuning nucleation kinetics via nonequilibrium chemical reactions},\ }\bibfield  {journal} {\bibinfo  {journal} {Physical Review Letters}\ }\textbf {\bibinfo {volume} {130}},\ \href {https://doi.org/10.1103/physrevlett.130.128203} {10.1103/physrevlett.130.128203} (\bibinfo {year} {2023})\BibitemShut {NoStop}%
\bibitem [{\citenamefont {Farag}\ \emph {et~al.}(2022)\citenamefont {Farag}, \citenamefont {Cohen}, \citenamefont {Borcherds}, \citenamefont {Bremer}, \citenamefont {Mittag},\ and\ \citenamefont {Pappu}}]{Farag}%
  \BibitemOpen
  \bibfield  {author} {\bibinfo {author} {\bibfnamefont {M.}~\bibnamefont {Farag}}, \bibinfo {author} {\bibfnamefont {S.~R.}\ \bibnamefont {Cohen}}, \bibinfo {author} {\bibfnamefont {W.~M.}\ \bibnamefont {Borcherds}}, \bibinfo {author} {\bibfnamefont {A.}~\bibnamefont {Bremer}}, \bibinfo {author} {\bibfnamefont {T.}~\bibnamefont {Mittag}},\ and\ \bibinfo {author} {\bibfnamefont {R.~V.}\ \bibnamefont {Pappu}},\ }\bibfield  {title} {\bibinfo {title} {Condensates of disordered proteins have small-world network structures and interfaces defined by expanded conformations},\ }\href@noop {} {\bibfield  {journal} {\bibinfo  {journal} {bioRxiv}\ ,\ \bibinfo {pages} {2022}} (\bibinfo {year} {2022})}\BibitemShut {NoStop}%
\bibitem [{\citenamefont {{B{\"o}ddeker}}\ \emph {et~al.}(2022)\citenamefont {{B{\"o}ddeker}}, \citenamefont {{Rosowski}}, \citenamefont {{Berchtold}}, \citenamefont {{Emmanouilidis}}, \citenamefont {{Han}}, \citenamefont {{Allain}}, \citenamefont {{Style}}, \citenamefont {{Pelkmans}},\ and\ \citenamefont {{Dufresne}}}]{Boeddeker}%
  \BibitemOpen
  \bibfield  {author} {\bibinfo {author} {\bibfnamefont {T.~J.}\ \bibnamefont {{B{\"o}ddeker}}}, \bibinfo {author} {\bibfnamefont {K.~A.}\ \bibnamefont {{Rosowski}}}, \bibinfo {author} {\bibfnamefont {D.}~\bibnamefont {{Berchtold}}}, \bibinfo {author} {\bibfnamefont {L.}~\bibnamefont {{Emmanouilidis}}}, \bibinfo {author} {\bibfnamefont {Y.}~\bibnamefont {{Han}}}, \bibinfo {author} {\bibfnamefont {F.~H.~T.}\ \bibnamefont {{Allain}}}, \bibinfo {author} {\bibfnamefont {R.~W.}\ \bibnamefont {{Style}}}, \bibinfo {author} {\bibfnamefont {L.}~\bibnamefont {{Pelkmans}}},\ and\ \bibinfo {author} {\bibfnamefont {E.~R.}\ \bibnamefont {{Dufresne}}},\ }\bibfield  {title} {\bibinfo {title} {{Non-specific adhesive forces between filaments and membraneless organelles}},\ }\href {https://doi.org/10.1038/s41567-022-01537-8} {\bibfield  {journal} {\bibinfo  {journal} {Nature Physics}\ }\textbf {\bibinfo {volume} {18}},\ \bibinfo {pages} {571} (\bibinfo {year} {2022})}\BibitemShut {NoStop}%
\bibitem [{\citenamefont {{Folkmann}}\ \emph {et~al.}(2021)\citenamefont {{Folkmann}}, \citenamefont {{Putnam}}, \citenamefont {{Lee}},\ and\ \citenamefont {{Seydoux}}}]{Folkmann_2021}%
  \BibitemOpen
  \bibfield  {author} {\bibinfo {author} {\bibfnamefont {A.~W.}\ \bibnamefont {{Folkmann}}}, \bibinfo {author} {\bibfnamefont {A.}~\bibnamefont {{Putnam}}}, \bibinfo {author} {\bibfnamefont {C.~F.}\ \bibnamefont {{Lee}}},\ and\ \bibinfo {author} {\bibfnamefont {G.}~\bibnamefont {{Seydoux}}},\ }\bibfield  {title} {\bibinfo {title} {{Regulation of biomolecular condensates by interfacial protein clusters}},\ }\href {https://doi.org/10.1126/science.abg7071} {\bibfield  {journal} {\bibinfo  {journal} {Science}\ }\textbf {\bibinfo {volume} {373}},\ \bibinfo {pages} {1218} (\bibinfo {year} {2021})}\BibitemShut {NoStop}%
\bibitem [{\citenamefont {Wu}\ \emph {et~al.}(2023)\citenamefont {Wu}, \citenamefont {King}, \citenamefont {Farag}, \citenamefont {Pappu},\ and\ \citenamefont {Lew}}]{wu2023single}%
  \BibitemOpen
  \bibfield  {author} {\bibinfo {author} {\bibfnamefont {T.}~\bibnamefont {Wu}}, \bibinfo {author} {\bibfnamefont {M.}~\bibnamefont {King}}, \bibinfo {author} {\bibfnamefont {M.}~\bibnamefont {Farag}}, \bibinfo {author} {\bibfnamefont {R.}~\bibnamefont {Pappu}},\ and\ \bibinfo {author} {\bibfnamefont {M.}~\bibnamefont {Lew}},\ }\bibfield  {title} {\bibinfo {title} {Single fluorogen imaging reveals spatial inhomogeneities within biomolecular condensates.},\ }\href@noop {} {\bibfield  {journal} {\bibinfo  {journal} {Biorxiv: the Preprint Server for Biology}\ } (\bibinfo {year} {2023})}\BibitemShut {NoStop}%
\bibitem [{\citenamefont {Lipiński}\ \emph {et~al.}(2022)\citenamefont {Lipiński}, \citenamefont {Visser}, \citenamefont {Robu}, \citenamefont {Fakhree}, \citenamefont {Lindhoud}, \citenamefont {Claessens},\ and\ \citenamefont {Spruijt}}]{Lipinski}%
  \BibitemOpen
  \bibfield  {author} {\bibinfo {author} {\bibfnamefont {W.~P.}\ \bibnamefont {Lipiński}}, \bibinfo {author} {\bibfnamefont {B.~S.}\ \bibnamefont {Visser}}, \bibinfo {author} {\bibfnamefont {I.}~\bibnamefont {Robu}}, \bibinfo {author} {\bibfnamefont {M.~A.~A.}\ \bibnamefont {Fakhree}}, \bibinfo {author} {\bibfnamefont {S.}~\bibnamefont {Lindhoud}}, \bibinfo {author} {\bibfnamefont {M.~M. A.~E.}\ \bibnamefont {Claessens}},\ and\ \bibinfo {author} {\bibfnamefont {E.}~\bibnamefont {Spruijt}},\ }\bibfield  {title} {\bibinfo {title} {Biomolecular condensates can both accelerate and suppress aggregation of alpha-synuclein},\ }\href {https://doi.org/10.1126/sciadv.abq6495} {\bibfield  {journal} {\bibinfo  {journal} {Science Advances}\ }\textbf {\bibinfo {volume} {8}},\ \bibinfo {pages} {eabq6495} (\bibinfo {year} {2022})}\BibitemShut {NoStop}%
\bibitem [{\citenamefont {Shen}\ \emph {et~al.}(2022)\citenamefont {Shen}, \citenamefont {Chen}, \citenamefont {Wang}, \citenamefont {Shen}, \citenamefont {Ruggeri}, \citenamefont {Aime}, \citenamefont {Wang}, \citenamefont {Qamar}, \citenamefont {Espinosa}, \citenamefont {Garaizar}, \citenamefont {George-Hyslop}, \citenamefont {Collepardo-Guevara}, \citenamefont {Weitz}, \citenamefont {Vigolo},\ and\ \citenamefont {Knowles}}]{ShenFUS}%
  \BibitemOpen
  \bibfield  {author} {\bibinfo {author} {\bibfnamefont {Y.}~\bibnamefont {Shen}}, \bibinfo {author} {\bibfnamefont {A.}~\bibnamefont {Chen}}, \bibinfo {author} {\bibfnamefont {W.}~\bibnamefont {Wang}}, \bibinfo {author} {\bibfnamefont {Y.}~\bibnamefont {Shen}}, \bibinfo {author} {\bibfnamefont {F.~S.}\ \bibnamefont {Ruggeri}}, \bibinfo {author} {\bibfnamefont {S.}~\bibnamefont {Aime}}, \bibinfo {author} {\bibfnamefont {Z.}~\bibnamefont {Wang}}, \bibinfo {author} {\bibfnamefont {S.}~\bibnamefont {Qamar}}, \bibinfo {author} {\bibfnamefont {J.~R.}\ \bibnamefont {Espinosa}}, \bibinfo {author} {\bibfnamefont {A.}~\bibnamefont {Garaizar}}, \bibinfo {author} {\bibfnamefont {P.~S.}\ \bibnamefont {George-Hyslop}}, \bibinfo {author} {\bibfnamefont {R.}~\bibnamefont {Collepardo-Guevara}}, \bibinfo {author} {\bibfnamefont {D.~A.}\ \bibnamefont {Weitz}}, \bibinfo {author} {\bibfnamefont {D.}~\bibnamefont {Vigolo}},\ and\ \bibinfo {author} {\bibfnamefont {T.~P.~J.}\ \bibnamefont {Knowles}},\ }\bibfield  {title} {\bibinfo
  {title} {Solid/liquid coexistence during aging of fus condensates},\ }\bibfield  {journal} {\bibinfo  {journal} {bioRxiv}\ }\href {https://doi.org/10.1101/2022.08.15.503964} {10.1101/2022.08.15.503964} (\bibinfo {year} {2022})\BibitemShut {NoStop}%
\bibitem [{\citenamefont {Toprakcioglu}\ \emph {et~al.}(2022)\citenamefont {Toprakcioglu}, \citenamefont {Kamada}, \citenamefont {Michaels}, \citenamefont {Xie}, \citenamefont {Krausser}, \citenamefont {Wei}, \citenamefont {Saric}, \citenamefont {Vendruscolo},\ and\ \citenamefont {Knowles}}]{KnowlesAggregatesSurface}%
  \BibitemOpen
  \bibfield  {author} {\bibinfo {author} {\bibfnamefont {Z.}~\bibnamefont {Toprakcioglu}}, \bibinfo {author} {\bibfnamefont {A.}~\bibnamefont {Kamada}}, \bibinfo {author} {\bibfnamefont {T.~C.~T.}\ \bibnamefont {Michaels}}, \bibinfo {author} {\bibfnamefont {M.}~\bibnamefont {Xie}}, \bibinfo {author} {\bibfnamefont {J.}~\bibnamefont {Krausser}}, \bibinfo {author} {\bibfnamefont {J.}~\bibnamefont {Wei}}, \bibinfo {author} {\bibfnamefont {A.}~\bibnamefont {Saric}}, \bibinfo {author} {\bibfnamefont {M.}~\bibnamefont {Vendruscolo}},\ and\ \bibinfo {author} {\bibfnamefont {T.~P.~J.}\ \bibnamefont {Knowles}},\ }\bibfield  {title} {\bibinfo {title} {Adsorption free energy predicts amyloid protein nucleation rates},\ }\href {https://doi.org/10.1073/pnas.2109718119} {\bibfield  {journal} {\bibinfo  {journal} {Proceedings of the National Academy of Sciences}\ }\textbf {\bibinfo {volume} {119}},\ \bibinfo {pages} {e2109718119} (\bibinfo {year} {2022})}\BibitemShut {NoStop}%
\bibitem [{\citenamefont {Laha}\ \emph {et~al.}(2024)\citenamefont {Laha}, \citenamefont {Bauermann}, \citenamefont {J{\"u}licher}, \citenamefont {Michaels},\ and\ \citenamefont {Weber}}]{laha}%
  \BibitemOpen
  \bibfield  {author} {\bibinfo {author} {\bibfnamefont {S.}~\bibnamefont {Laha}}, \bibinfo {author} {\bibfnamefont {J.}~\bibnamefont {Bauermann}}, \bibinfo {author} {\bibfnamefont {F.}~\bibnamefont {J{\"u}licher}}, \bibinfo {author} {\bibfnamefont {T.~C.}\ \bibnamefont {Michaels}},\ and\ \bibinfo {author} {\bibfnamefont {C.~A.}\ \bibnamefont {Weber}},\ }\bibfield  {title} {\bibinfo {title} {Chemical reactions regulated by phase-separated condensates},\ }\href@noop {} {\bibfield  {journal} {\bibinfo  {journal} {arXiv preprint arXiv:2403.05228}\ } (\bibinfo {year} {2024})}\BibitemShut {NoStop}%
\bibitem [{\citenamefont {{Uhl}}\ \emph {et~al.}(2021)\citenamefont {{Uhl}}, \citenamefont {{Weissmann}},\ and\ \citenamefont {{Seifert}}}]{BCsSeiffert}%
  \BibitemOpen
  \bibfield  {author} {\bibinfo {author} {\bibfnamefont {M.}~\bibnamefont {{Uhl}}}, \bibinfo {author} {\bibfnamefont {V.}~\bibnamefont {{Weissmann}}},\ and\ \bibinfo {author} {\bibfnamefont {U.}~\bibnamefont {{Seifert}}},\ }\bibfield  {title} {\bibinfo {title} {{Propagator for a driven Brownian particle in step potentials}},\ }\href {https://doi.org/10.1088/1751-8121/abc21f} {\bibfield  {journal} {\bibinfo  {journal} {Journal of Physics A Mathematical General}\ }\textbf {\bibinfo {volume} {54}},\ \bibinfo {eid} {065002} (\bibinfo {year} {2021})},\ \Eprint {https://arxiv.org/abs/2009.03201} {arXiv:2009.03201 [cond-mat.stat-mech]} \BibitemShut {NoStop}%
\bibitem [{\citenamefont {Piazza}(2022)}]{Piazza}%
  \BibitemOpen
  \bibfield  {author} {\bibinfo {author} {\bibfnamefont {F.}~\bibnamefont {Piazza}},\ }\bibfield  {title} {\bibinfo {title} {{The physics of boundary conditions in reaction–diffusion problems}},\ }\href {https://doi.org/10.1063/5.0128276} {\bibfield  {journal} {\bibinfo  {journal} {The Journal of Chemical Physics}\ }\textbf {\bibinfo {volume} {157}},\ \bibinfo {pages} {234110} (\bibinfo {year} {2022})}\BibitemShut {NoStop}%
\end{thebibliography}%
\clearpage

\appendix

\renewcommand{\thefigure}{S\arabic{figure}}
\setcounter{figure}{0} 

\section{Connecting a Fokker-Planck description to a free energy description}\label{sec: app FP to free energy}
In order to determine the conditional probability $P(\vec{x}_{/i},\vec{\sigma}_{/i}|{x_i,\sigma_i})$, we study the system close to equilibrium, so that all particles but the one in consideration are equilibrated. Then, we can connect the Fokker-Planck description of the problem to a statistical physics description \citep{Archer2004}. For better readability, in this section we denote $P(x_i, \sigma_i) = P_i$, $P_{/i} = P(\vec{x}_{/i},\vec{\sigma}_{/i}|x_i,\sigma_i)$ and $E_{i,\sigma_i} \equiv E(\vec{x}_{/i},x_i,\vec{\sigma}_{/i},\sigma_i)$. We write a general free energy as the sum of internal energy, entropy and a normalization factor for the conditional probability. 

\begin{equation} \label{eq: app gen free en}
\begin{split}
    F[P] =& 
    \sum_{\sigma_i}\int dx_i \SumInt_{/i} \Big[E_{i,\sigma_i}\cdot P_{/i} P_i + k_B T P_{/i} P_i \cdot \ln [P_{/i} P_i] \Big]
    \\&+ \lambda(x)(1- \SumInt_{/i} P_{/i})
\end{split}
\end{equation}
Taking the variation of this term with respect to $P_{/i}$, we find:
\begin{equation}
\begin{split}
&P_{/i} = \frac{e^{- \beta E_{i,\sigma_i} - 1 - \beta\frac{\lambda(x)}{P_i}}}{P_i}  
\end{split}
\end{equation}
which we can re-write using the normalization condition: 
\begin{equation} \label{eq: app p cond}
\begin{split}
P_{/i} = \frac{\exp \big[- \beta E_{i,\sigma_i}\big]}{Z(x_i,\sigma_i)}
\end{split}
\end{equation}
with the conditional partition function
\begin{equation}
\begin{split} \label{eq: app Z}
Z(x_i,\sigma_i) = \SumInt_{/i} e^{- \beta E_{i,\sigma_i}}
\end{split}
\end{equation}
This is the full partition function for the remaining $N-1$ particles. The corresponding conditional free energy is $F_{\text{cond}}(x_i,\sigma_i) = -k_B T \ln Z(x_i,\sigma_i)$. 

Plugging this expression in Eq.~\ref{eq: app gen free en}, the free energy becomes
\begin{eqnarray} \label{free energy for P}
F[P_i] &=& \sum_{\sigma_i}\int dx_i \Big[P_i F_{\text{cond}}(x_i, \sigma_i) +k_B T P_i \ln P_i 
\Big] \nonumber \\
&=& U[P_i] - T S[P_i] 
\end{eqnarray}
where the last expression highlights the natural split of the free energy into the internal energy and entropy for particle $i$.
We can use these results to re-express the diffusive part of Eq.~\ref{eq: marge}:
\begin{equation} 
    	\begin{split}
    	    \partial_t P_i &=  D~\nabla_i\big[\nabla_i P_i + \beta P_i \SumInt_{/i} \frac{e^{-\beta E_{i,\sigma_i}}}{Z(x_i,\sigma_i)} \nabla_i E_{i,\sigma_i}\big]\\&
    	    = D~\nabla_i\big[P_i \nabla_i \big(\ln(P_i) +  \ln(Z(x_i,\sigma_i))\big)\big] \\&
    	    = D~\nabla_i\big[P_i \nabla_i \frac{\delta F[P_i]}{\delta P_i}\big] = D~\nabla_i\big[ \nabla_i P_i + \nabla_i \frac{\delta U[P_i]}{\delta P_i}\big]
    	\end{split}
	\end{equation}

A full description of the system also requires us to specify $k^{eff}_{\sigma_i \sigma_i'}(x_i)$ as a function of thermodynamic quantities. As outlined in Section~\ref{sec: intro model}, we apply Kramers approximation, assuming that both the initial and the final points of the chemical reaction form local minima in the energy landscape along the reaction coordinate, separated by the energy barrier which we denote as $[\sigma_i\sigma_i']$. Then: 
\begin{equation} \label{eq: app kramer rates}
\begin{split} 
&k^{eff}_{\sigma_i\sigma_i'}(x_i) = k^0_{\sigma_i\sigma_i'} \SumInt_{/i} e^{-\beta E_{i,[\sigma_i\sigma_i']} + \beta E_{i,\sigma_i}} P_{/i}
\\&= k^0_{\sigma_i\sigma_i'} \exp\Big[-\beta\Big\{-\frac{1}{\beta}\ln\Big[\SumInt_{/i}  e^{-\beta E_{i,[\sigma_i\sigma_i']} + \beta E_{i,\sigma_i}} P_{/i}\Big]\Big\}\Big]
\end{split}
\end{equation}
Consistently with the previous approximation, we use again that all particles but the $i$th are in local equilibrium, and therefore we can express the conditional probability by Eq.~\ref{eq: app p cond}. Inserting it in Eq.~\ref{eq: app kramer rates}, the denominator of Eq.~\ref{eq: app p cond} cancels with one of the two energies from Kramer's rates. 
\begin{equation} \label{eq: app kramer rates 2}
\begin{split} 
k^{eff}_{\sigma_i\sigma_i'}(x_i) &= k^0_{\sigma_i\sigma_i'} \exp\Big[-\beta\Big\{-\frac{1}{\beta}\ln\Big[\SumInt_{/i}  \frac{e^{-\beta E_{i,[\sigma_i\sigma_i']}}}{Z(x_i,\sigma_i)}]\Big]\Big\}\Big] 
\end{split}
\end{equation}
By analogy with Eq.~\ref{eq: app p cond}, we can define the conditional  partition function of the energy barrier and the corresponding conditional barrier free energy $F_{\text{cond}}(x_i,[\sigma_i\sigma_i']) = -k_B T \ln Z(x_i,[\sigma_i\sigma_i'])$. With this, 
\begin{equation} \label{eq: app kramer rates 3}
k^{eff}_{\sigma_i\sigma_i'}(x_i) =  k^0_{\sigma_i\sigma_i'} \exp \left \{-\beta \left [F_{\text{cond}}(x_i,[\sigma_i\sigma_i']) -F_{\text{cond}}(x_i,\sigma_i)\right ] \right \} 
\end{equation}
Remarkably, the conditional free energies that modify the transition rates in Eq.\ref{eq: app kramer rates 3} can also be written as functional derivatives of the internal energy, just as the drift term, stressing the intimate connection between drift and chemical reactions.

Since the probability $P_i = P(x_i,\sigma_i)$ is the same for all particles in state $\sigma=\sigma_i$, we drop the index $i$ and, using the previous results, we can rewrite the full reaction-diffusion equation for the probability $P(x,\sigma)$ as:
\begin{eqnarray} \label{CH for P}
&& \partial_t P(x,\sigma) 
= D~\nabla \left [ P(x,\sigma) \nabla \frac{\delta F[P]}{\delta P(x,\sigma)} \right ] + \nonumber \\
&+& \sum_{\sigma '} \left [ k^{eff}_{\sigma' \sigma}(x) P(x, \sigma') - k^{eff}_{\sigma \sigma'}(x) P(x,\sigma) \right ]
\end{eqnarray}
where the usual form of the generalized Cahn-Hilliard equation for the diffusive part.

To further emphasize the connection with the generalized Cahn-Hilliard equation, it is possible to express Eq.\ref{CH for P} in terms of concentrations, by multiplying it by $N$ and using $c(x,\sigma) = N P(x,\sigma)$. We obtain
\begin{eqnarray} \label{CH for c}
&& \partial_t c(x,\sigma) 
= D~\nabla \left [ c(x,\sigma) \nabla \frac{\delta F[c]}{\delta c(x,\sigma)} \right ] + \nonumber \\
&+& \sum_{\sigma '} \left [ k^{eff}_{\sigma' \sigma}(x) c(x, \sigma') - k^{eff}_{\sigma \sigma'}(x) c(x,\sigma) \right ]
\end{eqnarray}
where we have used that, apart from an irrelevant additive constant, Eq.\ref{free energy for P} becomes
\begin{equation}
F[c] = \frac{1}{N}
\sum_{\sigma} \int dx
\Big[c(x,\sigma) F_{\text{cond}}(x, \sigma) + k_B T c(x,\sigma) \ln c(x,\sigma) 
\Big] \nonumber \\
\end{equation}
and that taking its functional derivative with respect to $P$ becomes, due to the $1/N$ term, the functional derivative with respect to $c$. Furthermore, also the conditional free energies in the exponent in Eq.\ref{eq: app kramer rates 3} can also be written as functional derivatives of the internal energy with respect to concentrations, making Eq.\ref{CH for c} only dependent on $c$.

With this result, we can rewrite Eq.~\ref{eq: marge} as Eq.~\ref{eq: 1-body FP}, as is done in Section~\ref{sec: intro model}.

\section{Derivation of no-flux-condition for two species out of equilibrium} \label{sec: app no flux 2 species}
For two species A and B, we can write the time evolution of their concentrations as:
\begin{equation*} 
    \begin{split}
        \partial_t c_A(x) =& D \partial_x [\partial_x c_A(x) + \beta c_A(x) \partial_x V_A(x)] \\&- k_{AB} c_A(x) + k_{BA} c_B(x) 
    \end{split}
\end{equation*}
\begin{equation}\label{one body prob}
    \begin{split}
        \partial_t c_B(x) =& D \partial_x [\partial_x c_B(x) + \beta c_B(x) \partial_x V_B(x)] \\&+ k_{AB} c_A(x) - k_{BA} c_B(x)
    \end{split}
\end{equation}
Solving the chemical part of the equation at steady state out of equilibrium, we get:
\begin{equation}
    \frac{c_A(x)}{c_B(x)} = \frac{k_{BA}(x)}{k_{AB}(x)} = \frac{k^0_{BA}}{k^0_{AB}}e^{-\beta(V_{A}(x)-V_{B}(x)+\mu_{A\rightarrow B})}
\end{equation}
We can express $c_B(x)$ as a function of $c_A(x)$:
\begin{equation}
    c_B(x) = \frac{k^0_{AB}}{k^0_{BA}}e^{\beta(V_{A}(x)-V_{B}(x)+\mu_{A\rightarrow B})}c_A(x)
\end{equation}
We can plug this into Eq.~\ref{one body prob} at steady state. The chemical part of the equation disappears by construction, so we are left only with the diffusive flux:
\begin{equation} 
    \begin{split}
        0 &= \frac{k^0_{AB}}{k^0_{BA}} e^{\beta(V_{A}(x)-V_{B}(x)+\mu_{A\rightarrow B})} \\&\cdot\partial_x [\partial_x  c_A(x) + \beta c_A(x) \partial_x V_{A}(x)]
    \end{split}
\end{equation}
Canceling $\frac{k^0_{AB}}{k^0_{BA}} e^{\beta(V_{A}(x)-V_{B}(x)+\mu_{A\rightarrow B})}$ on both sides, we recover the equation for $c_A(x)$. In this case, we can find a steady-state solution without fluxes for $c_A(x) = \hat{c}_A e^{-\beta V_A(x)}$, which determines also the solution for $c_B(x) = \hat{c}_A\frac{\hat{k}_{AB}}{\hat{k}_{BA}}e^{-\beta V_B(x)}$, where $\hat{c}_A$ can be found through the normalization. We find that the chemical and the diffusive part of Eq.~\ref{one body prob} are solved by the same probabilities and hence there are no diffusive fluxes for two species even when the system is out of equilibrium.

\section{Calculating the flux-less case for a tree system} \label{app: tree}

For a chemical system without cycles, every two concentrations of neighbouring nodes $\sigma$ and $\sigma'$ are connected via the detailed balance:

\begin{equation}
c_\sigma'(x) = \frac{k_{\sigma\sigma'}(x)}{k_{\sigma'\sigma}(x)} c_\sigma(x)
\end{equation}

Therefore, any two concentrations for states $\sigma$ and $\sigma'$ will be connected as

\begin{equation} \label{eq: app c ratio}
c_\sigma'(x) = K_{\sigma'\sigma}(x) c_\sigma(x)
\end{equation}

where $K_{\sigma'\sigma}(x)$ is the product of the rate ratios for all nodes along the path between $\sigma$ and $\sigma'$.

Substituting Eq.~\ref{eq: app c ratio} into Eq.~\ref{eq: 1-body FP} at steady state, we find that the chemical part is always satisfied as detailed balance was obeyed all along the path. Therefore, we are solving 

\begin{equation} \label{eq: app flux tree}
0 = \nabla\Big[\nabla c_\sigma'(x) +\beta c_\sigma'(x) \nabla F_{\sigma'}(x) )\Big]
\end{equation}

When we substitute Eq.~\ref{eq: app c ratio}, the first term becomes two: 
\begin{equation}\label{eq: app summands K c}
    \nabla K_{\sigma'\sigma}(x) \cdot c_\sigma(x) + K_{\sigma'\sigma}(x) \cdot \nabla c_\sigma(x)
\end{equation} According to Eq.~\ref{eq: k eff F}, the structure of $K_{\sigma'\sigma}(x)$ is the product of constant ratios of the relevant intrinsic transition rates and of the exponentials of the corresponding free energies. If the same state appears both as a final state for one reaction and an initial state for another, its energy will enter in the exponentials with opposite signs. Therefore, only $e^{-\beta(F_{\sigma'}(x)-F_\sigma(x))}$ will survive. Substituting this into the first term of Eq.~\ref{eq: app summands K c}, we recover $\beta K_{\sigma'\sigma}(x) c_\sigma(x) \nabla [F_\sigma(x,\sigma)-F_{\sigma'}(x)]$. Substituting this back into Eq.~\ref{eq: app flux tree}, we find that the resulting equation is identical to the equation for $c_\sigma(x)$. Therefore, for a tree-like reaction structure, the solution never exhibits fluxes as a consequence of the detailed balance for every chemical reaction.\\

\section{Calculating the flux-less case for a system with cycles} \label{app: cycle no-flux}

For a reaction network structure with cycles, we can use the Kirchhoff approach to express the concentration $c_\sigma(x)$ through the products of all rates along all edges $l$ leading to this state, summed over all possible spanning trees including it ($\mathcal{T}$) \citep{schnakenberg}.
\begin{equation*}
    c_\sigma(x) = \frac{\sum_{\lbrace \mathcal{T} \rbrace} {\prod_{l \in \mathcal{T}}} k_{l}^{\sigma}(x)}{\mathcal{N}(x)}
\end{equation*}
We can relate this concentration to another concentration, for example $c_1(x)$ via $c_\sigma(x) = \frac{c_\sigma(x)}{c_1(x)}c_1(x)$. Then,
\begin{equation}
\begin{split}
    c_\sigma(x) &= c_1(x) \frac{\sum_{\lbrace \mathcal{T} \rbrace} {\prod_{l \in \mathcal{T}}} k_{l}^{\sigma}(x)}{\sum_{\lbrace \mathcal{T} \rbrace} {\prod_{l \in \mathcal{T}}} k_{l}^{1}(x)}
    \\&= c_1(x) \frac{\sum_{\lbrace \mathcal{T} \rbrace} {\prod_{l \in \mathcal{T}}} \frac{k_{l}^{\sigma}(x)}{k_{l}^{\sigma}(x)}\prod_{l \in \mathcal{T}}k_l^1(x)}{\sum_{\lbrace \mathcal{T} \rbrace} {\prod_{l \in \mathcal{T}}} k_{l}^{1}(x)}
\end{split}
\end{equation}
Note that every path can be divided into the path $\gamma^\mathcal{T}_{\sigma1}$ that connects $\sigma$ and $1$ and sidebranches that are the same whether one is looking at all edges leading to $\sigma$ or to $1$. Therefore, for the ratio of rates we can replace the product over $l \in \mathcal{T}$ by one over $l \in \gamma^\mathcal{T}_{\sigma1}$. When substituting the Kramer form of $k_l^\sigma(x)$ we find analogously to before, $\prod_{l \in \gamma^\mathcal{T}_{\sigma1}} \frac{k_l^\sigma(x)}{k_l^1(x)} = e^{-\beta(F_\sigma(x)-F_1(x))}e^{-\beta \mu^l_{\sigma1}}$, where the free energy difference is independent of the taken path and $\mu^l_{\sigma1}$ is the total fuel chemical energy along the path. Therefore, we obtain
\begin{equation}\label{eq: app cycles conc}
\begin{split}
    c_\sigma(x) = c_1(x) e^{-\beta(F_\sigma(x)-F_1(x))} S_{1\sigma}(x)
\end{split}
\end{equation}
where we identified 
\begin{equation} \label{eq: app factor for cycles}
    S_{1\sigma}(x) = \frac{\sum_{\lbrace \mathcal{T} \rbrace} {\prod_{l \in \gamma^\mathcal{T}_{\sigma1}}} e^{-\beta \mu^l_{\sigma1}}\prod_{l \in \mathcal{T}}k_l^1(x)}{\sum_{\lbrace \mathcal{T} \rbrace} {\prod_{l \in \mathcal{T}}} k_{l}^{1}(x)}
\end{equation}
Substituting the Kramer form of $k_l^1(x)$, we note how every node except $1$ will appear as an initial state in $\prod_{l \in \mathcal{T}}k_{l}^{1}(x)$. Therefore, if we write $F_{\sigma'}(x) = F^b_{\sigma'}(x)+F^{init}_{\sigma'}(x)$, where $F^b_{\sigma'}(x)$ is the energy of the barrier in this direction along the edge and $F^{init}_{\sigma'}(x)$ is the energy of the initial state, then the product of all $F^{init}_{\sigma'}(x)$ will cancel and only the barrier energies will remain in Eq.~\ref{eq: app factor for cycles}:
\begin{equation} 
    S_{1\sigma}(x) = \frac{\sum_{\lbrace \mathcal{T} \rbrace} {\prod_{l \in \gamma^\mathcal{T}_{\sigma1}}} e^{-\beta \mu^l_{\sigma1}}\prod_{l \in \mathcal{T}}e^{-\beta F^b_l(x)}e^{-\beta \mu^l_1}}{\sum_{\lbrace \mathcal{T} \rbrace} {\prod_{l \in \mathcal{T}}} e^{-\beta F^b_l(x)}e^{-\beta \mu^l_1}}
\end{equation}
If the system is in equilibrium and all $\mu = 0$, $S_{1\sigma}(x)$ reduces to one and detailed balance is obeyed. Note that if the energies of the barriers, $F^b_l(x)$ are constant, even if offset from one another by a constant factor, $S_{1\sigma}(x)$ will be a constant, affecting the distribution, but resulting in a quasi-equilibrium state.\\
Now, plugging Eq.~\ref{eq: app cycles conc} into Eq.~\ref{eq: 1-body FP}, we obtain
\begin{equation}
\begin{split}
    \dot{c}_\sigma(x)=&\nabla[c_1(x)e^{-\beta(F_\sigma(x)-F_1(x))}S_{1\sigma}(x)]\\&+c_1 e^{-\beta(F_\sigma(x)-F_1(x))}S_{1\sigma}(x)\nabla F_\sigma(x) 
\end{split}
\end{equation}
which transforms into 
\begin{equation}
\begin{split}
    \dot{c}_\sigma(x)=&S_{1\sigma}(x)e^{-\beta(F_\sigma(x)-F_1(x))}\Big[\nabla c_1(x) + \beta c_1(x) \nabla F_1(x)\Big] \\&+ c_1(x) e^{-\beta(F_\sigma(x)-F_1(x))} \nabla S_{1\sigma}(x)
\end{split}
\end{equation}
The term in the square brackets is $\dot{c}_1(x)$, therefore $\dot{c}_1(x) = 0$ implies $\dot{c}_\sigma(x)=0$ only if $\nabla S_{1\sigma}(x) = 0$.

\section{Coarse-graining three states into two states with two pathways} \label{sec: app coarsegraining}
Consider the chemical cycle in Fig.~\ref{figmodel} in the case when C is a state that is so short-lived, so that its chemical reactions are much faster than its diffusion and it is always in steady state with respect to A and B. In that case, we can expess its concentration through the concentrations of the other two species,
\begin{equation}
c_C(x) = \frac{k_{AC}}{k_{CA}+k_{CB}}c_A(x) + \frac{k_{BC}}{k_{CA}+k_{CB}}c_B(x)
\end{equation}
We can substitute this solution in the equations for $c_A(x)$ and $c_B(x)$, which are not individually equilibrated:
\begin{equation}
\begin{split}
    \partial_t c_A &= -\partial_t c_B \\& = - (k_{AB}+\frac{k_{AC}k_{CB}}{k_{CA}+k_{CB}})c_A + (k_{BA}+\frac{k_{BC}k_{CA}}{k_{CA}+k_{CB}})c_B 
    \end{split}
\end{equation}
We can identify two pathways between A and B, where one corresponds to the rates $k_{AB}$ and $k_{BA}$, and the other -- to the rates 
\begin{equation}
\begin{split}
    &\Tilde{k}_{AB} = \frac{k_{AC}k_{CB}}{k_{CA}+k_{CB}}\\&
    \Tilde{k}_{BA} = \frac{k_{BC}k_{CA}}{k_{CA}+k_{CB}}
\end{split}
\end{equation}
Substituting the definitions of the rates, we find
\begin{equation}
\begin{split}
    \Tilde{k}_{AB} &= \frac{e^{-\beta(V_{AC}-V_A)}e^{-\beta(V_{BC}-V_C)}}{e^{-\beta(V_{AC}-V_C)}}+e^{-\beta(V_{BC}-V_C)}e^{\beta\mu_{A\rightarrow C}}\\&
    = e^{\beta V_A} \frac{e^{-\beta(V_{AC}+V_{CB}-V_C)}}{e^{-\beta(V_{AC}-V_C)}}+e^{-\beta(V_{BC}-V_C)}e^{\beta\mu_{A\rightarrow C}}
\end{split}
\end{equation}
where we can define the fraction in the last line to be $\exp(-\beta\Tilde{V}_{AB})$, the energy barrier of the second pathway. As this energy in general is not equal to $V_{AB}$, we indeed have two pathways that depend differently on the parameters of the system. \\
Note that this coarse-graining does not alter the affinity of the cycle:
\begin{equation}
    \frac{k_{AB}\Tilde{k}_{BA}}{k_{BA}\Tilde{k}_{AB}} = \frac{k_{AB}k_{BC}k_{CA}}{k_{BA}k_{AC}k_{CB}}
\end{equation}

Using this mapping, equation~\ref{eq: 1-body FP} can be solved analytically for a step potential (analogously to App.~\ref{app: BCs}):
\begin{equation}  \label{eq: 2 pathways}
\begin{split}
    &J_A = -J_B \\&=  
        \begin{cases}        
            \begin{aligned}\frac{e^{e_{R}+\epsilon_{L}}-e^{e_{L}+\epsilon_{R}}}{\mathcal{N}}(e^{\beta \Tilde{\mu}}-1)\sinh(q_R &L)\sinh(q_L(L+x)) \\& \mathrm{if}~-L \le x \le 0 \end{aligned} \\
            \begin{aligned}\frac{-e^{e_{R}+\epsilon_{L}}+e^{e_{L}+\epsilon_{R}}}{\mathcal{N}}(e^{\beta \Tilde{\mu}}-1)\sinh(q&_{L} L)\sinh(q_R(x-L)) \\& \mathrm{if}~0 \le x \le L \end{aligned}
        \end{cases}
\end{split}
\end{equation}
where $e_{L/R}$ and $\epsilon_{L/R}$ are the energy barriers along the two different chemical pathways and $\mathcal{N}$ is the overall normalisation factor. The factors  $q_L$ and $q_R$ are the square roots of the non-zero eigenvalues of the chemical transition matrix on the left and right side of the interface, respectively. The fluxes disappear of course in the absence of non-equilibrium driving ($\Tilde{\mu}=0$), but also when the energy barriers of the two pathways satisfy the condition $e_R-e_L = \epsilon_R-\epsilon_L$, which corresponds to barriers that do not depend on space or that do, but in the same way for both pathways. This setting could actually be re-mapped into a single pathway between A and B, which, as shown in Section \ref{sec: obtain fluxes}, does not create diffusive fluxes even away from equilibrium (see Appendix \ref{sec: app no flux 2 species}).

The spatial decay of the fluxes on each side of the interface is modulated only by the corresponding non-zero eigenvalue of the chemical transition matrix, which is (where $i \in \{L,R\}$): 
\begin{equation}
\begin{split}
q_i^2 =~& k_{AB}^0e^{-\beta e_i+\beta V_{A,i} + \beta\Tilde{\mu}} + k_{BA}^0e^{-\beta e_i+\beta V_{B,i}}
 + \\&
k_{AB}^0e^{-\beta\epsilon_i+\beta V_{A,i}} + k_{BA}^0e^{-\beta\epsilon_i+\beta V_{B,i}}
\end{split}
\end{equation}
This result provides the reaction-diffusion length 
\begin{equation}
l^{rd}_{L/R} = \sqrt{\frac{D}{q_{L/R}}} 
\end{equation}
For eigenvalues small enough that the argument of the hyperbolic sinus is small over the whole intervals $x \in [-L,0]$ and $x\in [0,L]$, the fluxes can be Taylor expanded to the first order, consistent with the observed linear behaviour.

\section{Numerical solution method} \label{sec: app num sol} 
In order to solve Eq.~\ref{eq: 1-body FP} numerically, we convert discretize space and solve the equation in every point. To do so, we construct a local stochastic matrix, the entries of which include both terms coming from the discretized diffusion and terms accounting for the chemical transformations. We then solve $\vec{0} = \mathbf{M}(x)\vec{\rho}(x)$, where $\vec{\rho}(x) = (\rho_A(1), \rho_A(2)...\rho_A(n), \rho_B(1), ... \rho_C(n))$ for a system discretized to $n$ spatial points. The steady-state solution of this equation is the non-zero eigenvector for the zero eigenvalue of the resulting stochastic matrix. We obtain it by dividing $\mathbf{M}$ into a reduced matrix $\mathbf{M}_{red}$ which lacks the first row and first column of $\mathbf{M}$ and into the vector $\vec{b}$, which is the first column of $\mathbf{M}$ without the first entry. We can choose the first entry of $\rho$, corresponding to the first (redundant) row of $\mathbf{M}$, freely and we choose it to be 1, so that $\vec{\rho} = (1,0,...,0)^T + \vec{\rho'}$. Then, we find that $0 = \mathbf{M} \rho = \vec{b}(1,0,...,0)^T+\mathbf{M}_{red}\vec{\rho'}$ and therefore $\vec{\rho'} = - \mathbf{M}_{red}^{-1} \vec{b}$. The final result is renormalized so that $\sum_{\sigma \in \{A,B,C\}}\sum_x \rho_\sigma(x) = 1$.\\
In the numerical solution, we can choose the form of the background potential to have an arbitrary functional form as long as it does not vary too much within one space unit.

\section{Boundary conditions for step potential} \label{app: BCs}

On each side, the system of equations for the three species can be represented by a stochastic matrix and written in the basis given by its eigenvectors $\vec{v}_i$. The concentration then can be written in the eigenvector basis, denoted as $\Tilde{c}(x)$. By construction, the matrix has one zero eigenvalue, $\lambda_0$. For it, we are solving $\partial_x \Tilde{c}_0(x) = 0$, while for the other two eigenvalues we solve $\partial_x^2 \Tilde{c}_i(x) = \lambda_i(x) \Tilde{c}_i(x)$, where $\lambda_i$ are the remaining two eigenvalues. In total, there are twelve free parameters and twelve boundary conditions. Of those, nine are coming from: 
\begin{itemize}
    \item the conservation of the total particle number (1 condition)
    \item no fluxes through the boundaries (at $x = -L$ and $x = L$) (6 conditions)
    \item the continuity of the fluxes at the interface (2 conditions)
\end{itemize}
The last three boundary conditions come from the behaviour of the particles immediately at the interface \citep[see also][]{BCsSeiffert, Piazza}. For a band around the interface with a width $\epsilon \rightarrow 0$, the contribution of chemical reactions is negligible and the dynamics of the particles is dominated by diffusion.  The probability of a particle crossing the interface is therefore proportional to the ratio of the Boltzmann factors on  either side. Intuitively, it can be seen as the rule to be obeyed by particles moving across the interface without undergoing chemical transitions.\\
The exact condition can be derived by integrating Eq.~\ref{eq: 1-body FP} around the interface. One finds:
\begin{equation}
\begin{split}
    0 &= D \int_{-\epsilon}^{\epsilon} \Big(\partial_x c_\alpha(x) + \beta c_\alpha(x) \partial_x V_\alpha(x) \Big) dx \\& 
    = D \int_{-\epsilon}^{\epsilon} e^{- \beta V_\alpha(x)}\partial_x (c_\alpha(x) e^{\beta V_\alpha(x)}) dx
\end{split}
\end{equation}
To fulfil this equation, $c_\alpha(x)$ has to have the form $c_\alpha(x) = C~\exp(-\beta V_\alpha(x))$ with a constant $C$. It follows that
\begin{equation}
    \frac{c_\alpha(0+)}{c_\alpha(0-)} = \frac{\exp(-\beta V_\alpha(0+))}{\exp(-\beta V_\alpha(0-))} 
\end{equation}
which provides the last three boundary conditions.

\section{Spherical coordinates} \label{sec: app spherical coords}
\begin{figure}
\centering
    \includegraphics[width=.4\paperwidth]{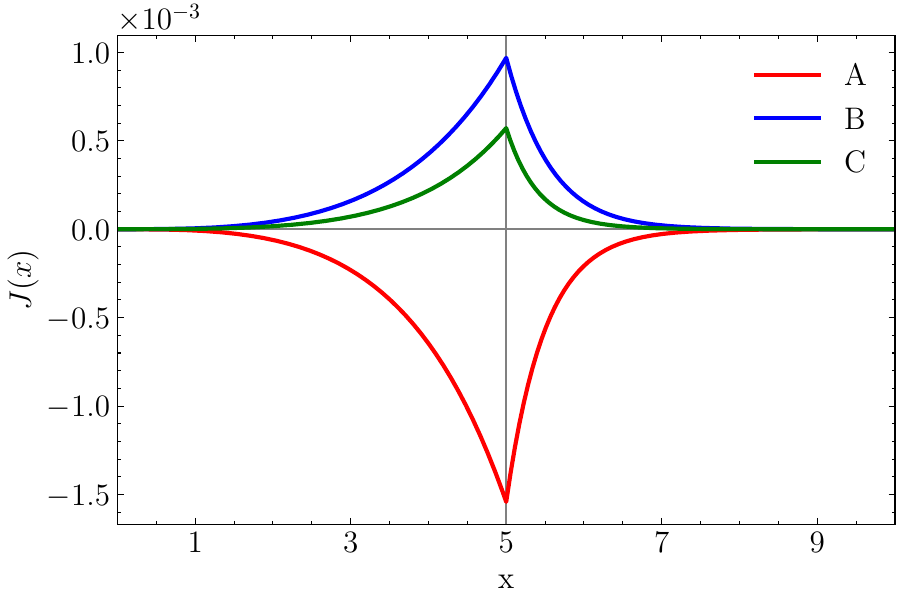}
\caption{Fluxes calculated for the step potential in 3D with the same parameters as in 1D. The system length is 10, the potential changes at $R = 5$.}
\label{Fig:stepfct3D}
\end{figure}

A condensate inside the cell is better described as a sphere in 3D. The fact that we are describing it in 1D does not alter the results qualitatively, as we show here for the solution of the problem for step potentials. We consider a spherical condensate centered around the origin with its interface at the radius R. The system is invariant under rotation, so that $c(r, \theta, \phi,\sigma) = c(r, \sigma)$. Then, we can solve the vectorial form of Eq.~\ref{eq: 1-body FP}
\begin{equation} \label{eq: 3D basic}
\partial_t \vec{c}(r) = 0 = D \vec{\nabla} (\vec{\nabla} \vec{c}(r) + \beta\vec{\nabla}\mathbf{V}(r)~\vec{c}(r)) + \mathbf{K}(r) \vec{c}(r) 
\end{equation}
where now the divergence is in spherical coordinates. We proceed as in Section~\ref{sec: max fluxes}, changing the basis of Eq.~\ref{eq: 3D basic} to the one spanned by the eigenvectors of the chemical transition matrix $\mathbf{K}(r)$ and solving the equation on the left and on the right of the interface, where the drift term is equal to zero. For the non-zero eigenvalues, the left-hand side becomes $\frac{1}{r^2}\partial_r(r^2\partial_r \Tilde{c}(r))$, so we look for a solution of the form of $\Tilde{c}(r) = f(r)/r$, for which we retrieve the same solutions as in the 1D case. Collecting all terms, the concentration profile inside the interface is described by
\begin{equation}
\begin{split}
    c^{in}(r) &= A^{in}_0 \vec{v}^{in}_0 + \left( \frac{A^{in}_{1+}}{r}e^{\lambda^{in}_{1} r} + \frac{A^{in}_{1-}}{r}e^{-\lambda^{in}_{1} r} \right) \vec{v}^{in}_1 \\&+ \left( \frac{A^{in}_{2+}}{r}e^{\lambda^{in}_{2} r} + \frac{A^{in}_{2-}}{r}e^{-\lambda^{in}_{2} r} \right) \vec{v}^{in}_2 
\end{split}
\end{equation}
and analogously outside of the interface. The boundary conditions will be analogous to the ones for 1D, with the difference that the no-flux conditions on the boundaries now transform into a no-flux condition at the outer boundary and a no-sources condition at the origin. Taking this into account, the solution can be determined as before. The result can be seen in Fig.~\ref{Fig:stepfct3D}. The fluxes still are the largest at the interface and the 1D results remain qualitatively unaltered. 

\clearpage

\end{document}